\documentclass[12pt]{elsarticle}                                          
\usepackage{graphicx}                                                         
\usepackage{a41}                                                                
\usepackage{xcolor}                     
\usepackage[rflt]{floatflt}
\usepackage{float}
\usepackage{lscape}
\usepackage{array}
\usepackage[english]{babel}
\usepackage[T1]{fontenc}
\usepackage{ae}
\usepackage{url}
\usepackage{amsmath, amsthm, amssymb}
\usepackage{slashed}

\usepackage{rotating}
\usepackage{graphicx}
\usepackage{comment}
\newcounter{mmacnt}
\def\restartmma{\setcounter{mmacnt}{0}}
\restartmma \catcode`|=\active
\def|#1|{\mathrm{#1}}
\catcode`|=12
\newenvironment{mma}{
\par\smallskip
\catcode`|=\active
\parskip=0pt\parindent=0pt 
\small
\def\In##1\\{%
\def\linebreak{\hfill\break\null\qquad}%
\refstepcounter{mmacnt}
\hangindent=2.5em\hangafter=0
\leavevmode
\llap{\tiny\sffamily In[\arabic{mmacnt}]:=\kern.5em}%
\mathversion{bold}\footnotesize$
\displaystyle##1$\normalsize
\mathversion{normal}\par
 }%
\def\Print##1\\{%
\def\linebreak{\hfill\break}%
\hangindent=2.5em\hangafter=0
\leavevmode ##1\par}%
\def\Out##1\\{%
\def\linebreak{$\hfill\break\null\hfill$}%
\kern\abovedisplayskip\par
\hangindent=2.5em\hangafter=0
\leavevmode
\llap{\tiny\sffamily Out[\arabic{mmacnt}]=\kern.5em}
\footnotesize$\displaystyle##1$
\normalsize\hfill\null\par
\kern\belowdisplayskip
}%
\def\Warning##1##2\\{%
\def\linebreak{\hfill\break}%
\hangindent=2.5em\hangafter=0
\leavevmode
{\scriptsize##1 : ##2}\par}%
}{%
\par\smallskip
}

\usepackage{color}
\newenvironment{fshaded}{%
\MakeFramed {\FrameRestore}
}%
{\endMakeFramed}

\makeatletter
\def\ps@pprintTitle{%
\let\@oddhead\@empty
\let\@evenhead\@empty
\def\@oddfoot{\reset@font\hfil\thepage\hfil}
\let\@evenfoot\@oddfoot
}
\makeatother
\usepackage{tikz}
\usetikzlibrary{matrix}
\allowdisplaybreaks[4]

\begin{document} 
\begin{frontmatter}
\title{\Large
\textbf{
Probing dark matter through charged Higgs 
pair production at future multi-TeV 
muon colliders: A machine-learning analysis
}}
\author[1]{Khiem Hong Phan}
\ead{khiem.phan@eiu.edu.vn}
\author[2]{Quang Hoang-Minh Pham}
\address[1]{\it Eastern International University, 
81 Nam Ky Khoi Nghia Street, Binh Duong Ward, Ho Chi Minh City, Vietnam}
\address[2]
{\it VNUHCM-University of Science,
$227$ Nguyen Van Cu, District $5$,
Ho Chi Minh City $70000$, Vietnam}
\pagestyle{myheadings}
\markright{}
\begin{abstract} 
Probing dark matter (DM) via charged Higgs pair production
at future multi-TeV muon colliders is investigated within
the Inert Doublet Model (IDM). The viable parameter space
of the IDM is first updated by incorporating theoretical
constraints and current experimental data. Based on the
allowed parameter space, we evaluate DM relic density and
compare the results with the latest constraints from direct
DM detection experiments. The resulting parameter points
consistent with all DM constraints are subsequently employed
to study charged Higgs pair production at future multi-TeV
muon colliders, including the subsequent decays of the
charged Higgs bosons into Standard Model (SM) particles
in association with DM candidate. In particular, we study
the following production processes:
$\mu^- \mu^+ \to \nu_{\mu} \bar{\nu}_{\mu}
H^{\pm} H^{\mp}
\to \ell^+ \ell^-
+
\nu_{\mu} \bar{\nu}_{\mu}
\nu_{\ell} \bar{\nu}_{\ell} HH$,
$\mu^- \mu^+ \to \nu_{\mu} \bar{\nu}_{\mu}
H^{\pm} H^{\mp}
\to \ell^\pm + 2\,\text{jets}
+ \nu_{\mu} \bar{\nu}_{\mu}
\nu_{\ell} HH,
$
and
$
\mu^- \mu^+ \to \nu_{\mu} \bar{\nu}_{\mu}
H^{\pm} H^{\mp}
\to 4\,\text{jets}
+ \nu_{\mu} \bar{\nu}_{\mu} HH$
for $\ell =e, \mu$.
The signal significance is evaluated
against the corresponding SM backgrounds
using both cut-based and machine-learning
(ML) approaches. We find that ML framework 
substantially enhances the sensitivity 
to the signal
processes compared with the conventional
cut-based analysis. Furthermore,
our results indicate that the DM
signals through charged Higgs
pair production can be indirectly probed
with a statistical significance exceeding
$5\sigma$ for several viable benchmark
points at future multi-TeV muon colliders.
\end{abstract}
\begin{keyword} 
\footnotesize
Higgs boson phenomenology,
DM searches at future 
colliders,
new physics searches at current 
and future colliders,
physics beyond
the Standard Models.
\end{keyword}
\end{frontmatter}
\section{Introduction}
Despite its remarkable success, 
the SM of particle physics fails 
to explain several fundamental 
observations, including the existence 
of the DM, the tiny neutrino masses, 
and the baryon asymmetry of the Universe, etc.
These shortcomings of the SM strongly 
suggest the existence of physics beyond 
the Standard Model (BSM).
Furthermore, although the discovery of the SM-like
Higgs boson at the Large Hadron
Collider (LHC)~\cite{ATLAS:2012yve,CMS:2012qbp}
confirmed the mechanism of electroweak symmetry
breaking (EWSB), the SM provides no fundamental
theoretical principle that determines
the structure of its Higgs scalar sector.
Since many open questions in particle physics
are closely connected to this sector, a wide
variety of BSM frameworks  introduce
extended scalar sectors.
In addition, the Planck
Collaboration~\cite{Planck:2018nkj,Planck:2018vyg}
has measured
the DM relic density with high precision. 
Despite significant experimental and observational
progress, the nature of DM remains one of the
most important unresolved problems in modern physics.
Weakly interacting massive particles (WIMPs)
represent one of the most well-motivated DM
candidates, as they naturally arise at the
electroweak scale and can be probed at collider
experiments. 
In this context, extensions of
the Higgs scalar sector provide an attractive
framework that can simultaneously address
several shortcomings of the SM while offering
viable DM candidates.
Among various BSM scenarios,
the IDM represents one of the simplest
extensions of the SM by introducing
an additional scalar doublet. An exact
$Z_2$ symmetry is imposed on the IDM
Lagrangian, under which all SM fields
are even, while the additional scalar
doublet is odd. This doublet contains
two neutral scalars $H$ and $A$, 
together
with a pair of charged scalars $H^\pm$.
As a consequence of the conserved
$Z_2$ symmetry, these new scalar particles
interact with SM fields only through
gauge interactions and scalar
self-interactions. The lightest
$Z_2$-odd scalar is stable and can
therefore serve as a viable DM candidate. 
Moreover, when the model
is extended to include right-handed
neutrinos, the observed smallness of
neutrino masses can be naturally generated
through the radiative seesaw 
mechanism~\cite{Tao:1996vb, Ma:2006km}. 

Precision measurements of the Higgs scalar
sector constitute one of the central
objectives of future collider experiments,
as the resulting data may provide profound
insights into the underlying nature of
EWSB and offer stringent tests of various
BSM scenarios. In this framework, extensive 
phenomenological investigations of 
scalar particle production, together with 
their implications for DM physics, have been 
carried out in the literature.
In particular, production mechanisms of the
additional scalar states predicted by the
IDM as well as implication of 
DM have been widely explored at both hadron 
and lepton colliders.
At hadron colliders, IDM scalar production 
was investigated in Ref.~\cite{Ghosh:2021noq}. 
DM signatures within the IDM framework were explored at the LHC in Ref.~\cite{Belyaev:2016lok}, while searches based on vector-boson-fusion (VBF) processes were examined in Refs.~\cite{Dercks:2018wch,Dutta:2017lny}. 
Collider probes of the decay channel $H^\pm \to HW^\pm$ through dijet plus missing-energy signatures were proposed in Ref.~\cite{Diaz:2015pyv}. Additional studies at the LHC addressed mono-$W$ signatures~\cite{Wan:2018eaz}, dijet plus missing transverse energy final states~\cite{Poulose:2016lvz}, multilepton signatures~\cite{Datta:2016nfz,Gustafsson:2012aj}, and indirect probes of IDM scalar effects through Higgs-strahlung processes~\cite{He:2024bwh}. Ref.~\cite{Fan:2022dck} further examined the IDM in connection with DM phenomenology, the CDF II $W$-boson mass anomaly, and future detection prospects.
At future lepton colliders, charged scalar pair production at high-energy CLIC was studied in Refs.~\cite{Klamka:2021vqp,Klamka:2022ukx}, while scalar-pair production at $e^+e^-$ colliders was investigated in Refs.~\cite{Kalinowski:2018ylg,Hashemi:2015swh}. Collider signatures of inert scalar bosons at the ILC were discussed in Ref.~\cite{Aoki:2013lhm}. The process $e^+e^- \to HA$ in a linearly polarized laser field was analyzed in Ref.~\cite{Ouhammou:2023llz}, whereas DM searches through charged Higgs pair production at a $\gamma\gamma$ collider were proposed in Ref.~\cite{Guo-He:2020nok}. More recently, the potential of muon colliders to probe the IDM via VBF processes was explored in Ref.~\cite{Braathen:2024ckk}.
The IDM has also been extensively studied from the perspective of DM phenomenology. Early analyses of the IDM following the discovery of the SM-like Higgs boson at the LHC were presented in Ref.~\cite{Goudelis:2013uca}. Fully automated computations of loop-induced DM annihilation processes relevant for indirect-detection observables were subsequently reported in Ref.~\cite{Arina:2021gfn}. Monochromatic gamma-ray signatures from scalar DM annihilation, together with the associated antimatter cosmic-ray signals, were investigated in Refs.~\cite{Gustafsson:2007pc,Nezri:2009jd}. The dominant electroweak corrections to the DM direct-detection cross section arising from one-loop gauge-boson-mediated diagrams were analyzed in Ref.~\cite{Klasen:2013btp}.
Beyond conventional cut-based analyses, advanced collider-search strategies employing multivariate techniques and jet-substructure observables were investigated in Ref.~\cite{Bhardwaj:2019mts}. Comprehensive global studies of the IDM, incorporating collider observables together with direct and indirect DM constraints, were performed in Ref.~\cite{Eiteneuer:2017hoh}.
More generally, one-loop radiative corrections within the investigated model have played an important role in improving the precision of theoretical predictions.
One-loop 
corrections to the triple-Higgs coupling and associated Higgs production processes in the IDM were examined 
in Refs.~\cite{Arhrib:2015hoa,Arhrib:2014pva}. Furthermore,
one-loop radiative corrections to the trilinear Higgs self-coupling were computed in Ref.~\cite{Falaki:2023tyd}. In addition, complete one-loop electroweak corrections to the processes $e^+e^- \to Zh^0/H^0A^0$ and $e^+e^- \to H^+H^-$ were evaluated in Refs.~\cite{Abouabid:2020eik,Abouabid:2022rnd}. One-loop QED and
weak corrections to
${\gamma}{\gamma} \to H^-H^+$ were also studied
in Ref.~\cite{Abouabid:2025whn}.

Future lepton colliders~\cite{Apollinari:2017lan,FCC:2018evy,InternationalMuonCollider:2025sys}, which are expected to provide cleaner environments than hadron colliders, have been proposed as promising facilities for probing the Higgs scalar sector with high precision. Precision measurements at such colliders may also offer indirect sensitivity to DM through deviations from SM predictions. In this work,
we investigate DM signatures through charged Higgs pair production at
future colliders within the framework of
the IDM. The viable parameter space of 
the model is first identified by imposing 
both theoretical constraints and current
experimental bounds. We subsequently
evaluate the DM relic density, and only
parameter points consistent with present
cosmological observations are retained
for the phenomenological analysis.
Particular emphasis is placed on probing 
DM via charged
Higgs boson pair production at future
multi-TeV muon colliders.
The following signal channels are
considered:
$
\mu^- \mu^+ \to
\nu_{\mu} \bar{\nu}_{\mu}
H^{\pm} H^{\mp}
\to
\ell^+ \ell^-
+
\nu_{\mu} \bar{\nu}_{\mu}
\nu_{\ell} \bar{\nu}_{\ell} HH,
$
$
\mu^- \mu^+ \to
\nu_{\mu} \bar{\nu}_{\mu}
H^{\pm} H^{\mp}
\to
\ell^\pm + 2\,\mathrm{jets}
+ \nu_{\mu} \bar{\nu}_{\mu}
\nu_{\ell} HH,
$
and
$
\mu^- \mu^+ \to
\nu_{\mu} \bar{\nu}_{\mu}
H^{\pm} H^{\mp}
\to
4\,\mathrm{jets}
+ \nu_{\mu} \bar{\nu}_{\mu} HH
$ for $\ell = e, \mu$.
Our numerical results demonstrate that
DM signals via charged Higgs boson 
pairs can be
efficiently probed indirectly at future
high-energy muon colliders, thereby
highlighting the strong potential of
such facilities for exploring the IDM
parameter space and unveiling possible
signatures of BSMs.

The organization of this
paper is as follows. In Sect.~3,
we briefly review the IDM and
summarize the relevant parameter
space. Section~3 is devoted to
probe DM via charged
Higgs pair production at future
multi-TeV muon colliders, where
the corresponding phenomenological
results are presented. Finally,
conclusion is devoted in Sect.~4.
\section{Review of the Inert Doublet Model 
and its constrainst}
In this section, we briefly review
the theoretical framework of the
IDM under consideration and summarize
the relevant theoretical constraints
together with the latest experimental
constraints on the model parameter space.
\subsection{A brief 
description of the IDM}
Compared with the SM, the IDM contains an
additional $SU(2)_L$ scalar doublet. The scalar 
sector therefore consists of two $SU(2)_L$ scalar 
doublets with hypercharge $Y=1$, denoted by 
$\Phi_1$ and $\Phi_2$.
The model is invariant under an
exact discrete $Z_2$ symmetry,
under which $\Phi_2 \to -\Phi_2$,
while $\Phi_1$ and SM fields 
are $Z_2$-even. This symmetry 
forbids Yukawa
couplings of $\Phi_2$ to fermions
for ensuring the stability of the
lightest $Z_2$-odd particle,
which can serve as a viable dark
matter candidate. The most general
renormalizable scalar potential
consistent with the
$SU(2)_L \times U(1)_Y$ gauge
symmetry and the $Z_2$ symmetry
is given by~\cite{Braathen:2024ckk}
\begin{eqnarray}
\mathcal{V}(\Phi_1,\Phi_2) 
&=& \mu_1^2 |\Phi_1|^2 
+ \mu_2^2 |\Phi_2|^2
+ \frac{\lambda_1}{2} 
|\Phi_1|^4 
+
\frac{\lambda_2}{2}
|\Phi_2|^4  
+ \lambda_3 |\Phi_1|^2 |\Phi_2|^2
+ \lambda_4 |\Phi_1^\dagger \Phi_2|^2
\nonumber \\
&&
+ \frac{\lambda_5}{2}
\left[(\Phi_1^\dagger \Phi_2)^2 
+ \mathrm{h.c.}\right].
\end{eqnarray}
After EWSB, both scalar fields
$\Phi_1$ and $\Phi_2$ are
parametrized as follows
\begin{eqnarray}
\Phi_1 &=&
\begin{bmatrix}
G^+ \\
(v+ h +iG_0)/\sqrt{2}
\end{bmatrix}
, \; 
\Phi_2 =
\begin{bmatrix}
H^+ \\
(H +iA)/\sqrt{2}
\end{bmatrix}.
\label{representa-htm}
\end{eqnarray}
The scalar doublet $\Phi_1$ only develops a
vacuum expectation value (VEV)
$v \simeq 246~\mathrm{GeV}$, whereas $\Phi_2$
does not develop a VEV, thereby preserving
the discrete $Z_2$ symmetry.
The physical scalar spectrum of the model
contains the SM-like Higgs boson $h$, which
originates from $\Phi_1$ and was observed
at the LHC, together with four additional
inert scalar states arising from $\Phi_2$,
namely a pair of charged scalars $H^\pm$
and two neutral scalars $H$ and $A$.
Depending on the choice of model parameters,
the lightest neutral inert scalar can serve
as a viable dark matter candidate. In this
work, we assume $\lambda_5$ to be real and
negative, such that the scalar $H$ corresponds
to the DM candidate. The masses of
the physical inert scalar states are given by
\begin{eqnarray}
m_h^2 &=& -2\mu_1^2 
= 2\lambda_1 v^2,\\
m_H^2 &=& \mu_2^2 
+ v^2\lambda_L,\\
m_{A}^2 &=& \mu_2^2 
+ v^2\lambda_R,\\
m_{H^\pm}^2 &=& \mu_2^2 
+ \dfrac{v^2}{2}\lambda_3.
\end{eqnarray}
Where $\lambda_{L/R} 
= (\lambda_3 + \lambda_4 \pm \lambda_5)/2$. 
The bare parameters of the scalar
potential can be expressed in terms
of the physical scalar masses by
solving the above system of equations,
leading to the following relations:
\begin{eqnarray}
\lambda_1 &=& \frac{m_{h}^2}{2v^2}, \\
\lambda_3 &=& 
\frac{2(m_{H^\pm}^2-\mu_2^2)}{v^2}, \\
\lambda_4 &=& \frac{m_{H}^2 
+m_{A}^2-2m_{H^\pm}^2}{v^2},        \\
\lambda_5 &=& \frac{m_{H}^2
-m_{A}^2}{v^2}.
\end{eqnarray}
In this work, the parameter space of the IDM
is characterized by the following independent
parameters: the inert scalar masses
$m_H$, $m_A$, and $m_{H^\pm}$, together with
the quartic couplings $\lambda_2$ and
$\lambda_L$. A useful relation frequently
employed throughout our analysis is given by
\begin{eqnarray}
\lambda_R = \lambda_L 
+ \dfrac{m_A^2 - m_H^2}{v^2}.
\end{eqnarray}
The triple and quartic Higgs
couplings relevant to our analysis
are given below:
\begin{eqnarray}
h H H &=& -2v\lambda_L 
= -v(\lambda_3+\lambda_4+\lambda_5) 
= v\lambda_{hHH}, 
\\
h A A &=& -2v\lambda_R 
= -v(\lambda_3+\lambda_4-\lambda_5) 
= v\lambda_{hAA}, 
\\
h H^\pm H^\mp &=& -v\lambda_3 
= v\lambda_{hH^\pm H^\mp},
\\
H H A A &=& -2\lambda_2.
\end{eqnarray}
In the phenomenological analysis presented
in the next section, the considered
processes are particularly sensitive to
the $h H^\pm H^\mp$ coupling, whose
strength can be significantly enhanced
in regions of parameter space associated
with large values of $\lambda_3$.
Before presenting to the phenomenological
analysis of DM signatures in the
IDM through charged Higgs pair production
at future multi-TeV muon colliders, we
first discuss the relevant theoretical and
experimental constraints imposed on the
IDM parameter space.
\subsection{The current 
constraints}
The viable parameter space of the IDM,
subject to both theoretical and
experimental constraints, is presented
in this subsection. For comprehensive
discussions of these constraints, we
refer the reader to
Refs.~\cite{Abouabid:2023cdz,
Abouabid:2025bpm,Aiko:2023nqj}.
The theoretical constraints originate
from the inert vacuum condition, the
bounded-from-below requirement of the
scalar potential, and perturbative
unitarity. These theoretical conditions
are briefly summarized as follows:
\begin{itemize}
\item {\bf Inert vacuum condition:}
This condition guarantees that the
scalar doublet $\Phi_2$ remains inert,
namely that it does not develop a VEV.
At leading order (LO), the following
condition must be satisfied
~\cite{Ginzburg:2010wa}:
\begin{eqnarray}
\dfrac{\mu_2^2}{\sqrt{\lambda_2}}
\geq 
\dfrac{\mu_1^2}{\sqrt{\lambda_1}}.
\end{eqnarray}
\item {\bf Vacuum stability condition:}
This condition ensures that the scalar
potential remains bounded from below
in the large-field limit. The
necessary requirements for vacuum
stability are expressed as
follows~\cite{Kanemura:1999xf}:
\begin{eqnarray}
 \lambda_1 \geq 0,\quad
 \lambda_2 \geq 0, \quad
 \sqrt{\lambda_1 \lambda_2}
 + \lambda_3 + \textrm{min}
 \{0,\lambda_4-\lambda_5\} 
 \geq 0.
\end{eqnarray}
It should also be noted that
$\mu_1^2 < 0$ follows from the
stationary condition, whereas
$\mu_2^2 > 0$ is necessary for
the stability of the inert vacuum.
\item {\bf Perturbative unitarity:}
This constraint requires
$|a_0^j| < 1/2$, where $a_0^j$
denotes the eigenvalues of the
$s$-wave scattering amplitude matrix.
The two-to-two scalar-boson scattering
processes are taken into account in
the high-energy limit. We refer the
reader to
Refs.~\cite{Kanemura:1993hm, Akeroyd:2000wc}
for the derivation of 
the explicit expressions of
$a_0^j$.
In addition, we impose a further
theoretical constraint requiring all
quartic couplings in the scalar
potential to satisfy the perturbativity
condition $|\lambda_i| < 8\pi$,
thereby ensuring the perturbative
validity of the model under
investigation.
More explicitly, the following
relations must be fulfilled:
\begin{eqnarray}
|e_i| \le 8\pi, \qquad i = 1,\ldots,12.
\end{eqnarray}
The explicit expressions for $e_i$
are given by~\cite{Kanemura:1993hm, Akeroyd:2000wc}
\begin{eqnarray}
e_{1,2} &=& \lambda_3 \pm \lambda_4, \\
e_{3,4} &=& \lambda_3 \pm \lambda_5, \\
e_{5,6} &=& \lambda_3 + 2\lambda_4
\pm 3\lambda_5,
\\
e_{7,8} &=& -\lambda_1 - \lambda_2
\pm \sqrt{(\lambda_1-\lambda_2)^2
+ \lambda_5^2}, \\
e_{9,10} &=& -3\lambda_1 - 3\lambda_2
\pm \sqrt{9(\lambda_1-\lambda_2)^2
+ (2\lambda_3+\lambda_4)^2}, \\
e_{11,12} &=& -\lambda_1 - \lambda_2
\pm \sqrt{(\lambda_1-\lambda_2)^2
+ \lambda_4^2}.
\end{eqnarray}
\end{itemize}
We next turn to the experimental
constraints, which are imposed through
the following conditions.
First, constraints arising from the
precise measurements of the gauge-boson
widths and electroweak precision
observables (EWPOs) at LEP-II and the
LHC are taken into account. These
measurements imply that additional
vector-boson decay channels predicted
in the IDM, such as
$W^\pm \to H H^\pm,\, A H^\pm$
and
$Z \to H A,\, H^\pm H^\mp$,
must not lead to sizable deviations
from the experimentally measured
widths. Consequently, the kinematic
closure of these decay modes requires
the following conditions
\cite{Goudelis:2013uca,Abouabid:2020eik,
Lundstrom:2008ai,
Swiezewska:2012eh,Arhrib:2012ia}:
\begin{eqnarray}
2m_{H^\pm}\geq m_Z,\quad
 m_H+m_{H^\pm}\geq m_W, \quad
 m_H+m_A\geq m_Z,  \quad
 m_A+m_{H^\pm}\geq m_W.
\end{eqnarray}
Furthermore, searches for
$e^- e^+ \to H^\pm H^\mp$
at LEP-II~\cite{Pierce:2007ut}
impose the lower bound
$m_{H^\pm} \gtrsim 90~\mathrm{GeV}$.
On the other hand, searches for
$AH$ production reported in
Ref.~\cite{Lundstrom:2008ai}
exclude the regions
$m_H < 80~\mathrm{GeV}$
and
$m_A < 100~\mathrm{GeV}$.
In addition, EWPO constraints impose
important restrictions on the inert
scalar spectrum, particularly on the
mass splittings among the inert states,
since these particles contribute to
the gauge-boson self-energies through
loop corrections. Such effects are
conveniently parameterized in terms
of the oblique parameters
$S$, $T$, and $U$~\cite{Hessenberger:2016atw}.
In particular, the experimental values of 
the oblique parameters are taken to be
$S_0= -0.05 \pm 0.07$, $T_0 = 0.0\pm0.06$,
and $\rho_{ST}=0.93$.
For all the constraints discussed above,
we employ the public package
{\tt 2HDMC}~\cite{Eriksson:2009ws}.
It should also be emphasized that
constraints from Higgs precision
measurements at the LHC, together with
bounds from direct scalar searches at
colliders, are evaluated using the
public tools
{\tt 2HDMC}~\cite{Eriksson:2009ws}
and
{\tt HiggsBounds}~\cite{Bahl:2022igd}.

Another important observable constraining
the IDM parameter space is the trilinear
self-coupling of the SM-like Higgs boson,
$\lambda_{hhh}$. The one-loop corrections
to $\lambda_{hhh}$ in BSM scenarios
are evaluated using the public package
{\tt anyH3}~\cite{Bahl:2023eau}. At present,
the strongest constraints on this coupling
come from searches for Higgs boson pair
production at the LHC. The experimental
upper limits on the di-Higgs production
cross section can be interpreted as
constraints on the modified Higgs
self-coupling, defined by~\cite{Bahl:2023eau}
\begin{eqnarray}
\kappa_{hhh} = 
\dfrac{\lambda_{hhh}^{\textrm{IDM} }}
{\lambda_{hhh}^{\textrm{SM}}}. 
\end{eqnarray}
The observed $95\%$ confidence level
(CL) limits on $\kappa_{hhh}$ reported
by the ATLAS Collaboration are given by
$\kappa_{hhh}\in[-1.4,\,6.1]$ as 
in~\cite{ATLAS:2022jtk},
whereas the CMS Collaboration obtained
the range
$\kappa_{hhh}\in[-1.4,\,7.8]$
as shown in~\cite{CMS:2024awa}.
Moreover, the combined ATLAS and CMS
analysis constrains the modified Higgs
self-coupling to
$\kappa_{hhh}\in[-1.2,\,7.2]$
at the $95\%$ CL~\cite{ATLAS:2024ish}.
These combined bounds are adopted
throughout the present analysis.
We also emphasize that the presence
of DM in the IDM can induce
an additional invisible decay channel
for the SM-like Higgs boson.
Measurements performed by the ATLAS
Collaboration place a stringent upper
bound on the invisible Higgs branching
ratio,
$\mathrm{Br}(h\to \text{invisible})
\leq 10.7\%$.
This constraint, in turn, imposes a
lower bound on the DM mass,
$m_H \gtrsim 40~\mathrm{GeV}$,
as reported in
Ref.~\cite{ATLAS:2023tkt}.

\begin{figure}[H]
\centering
\begin{tabular}{cc}
\includegraphics[width=8.5cm, height=8cm]
{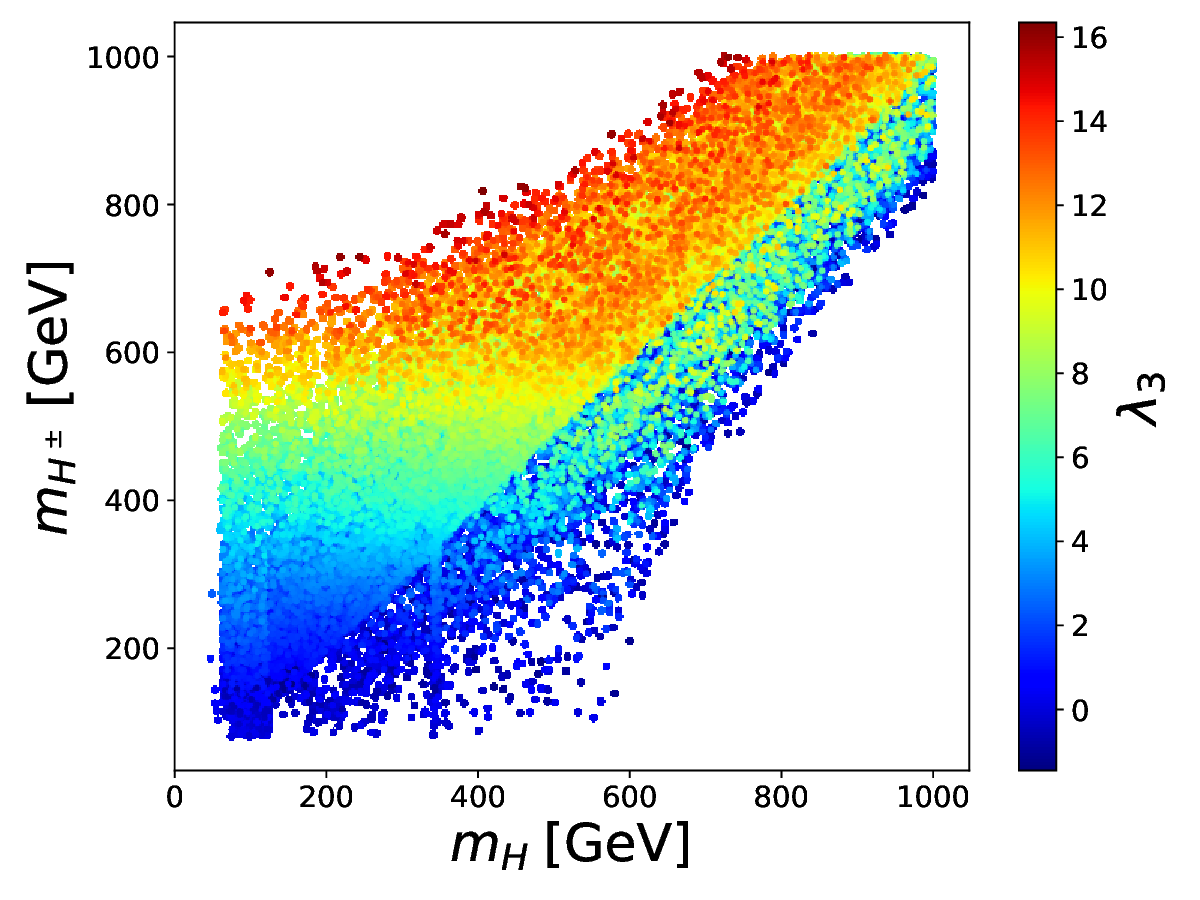}
&
\includegraphics[width=8.5cm, height=8cm]
{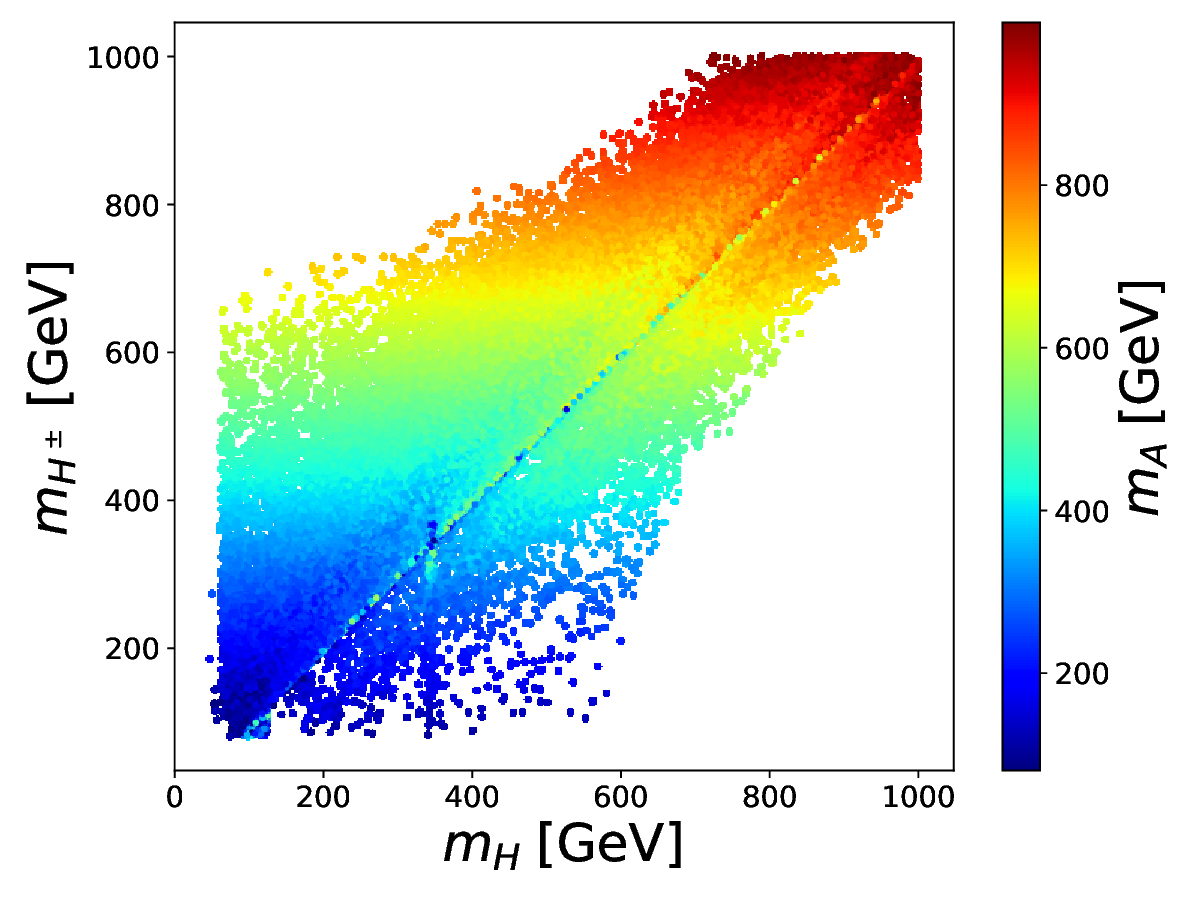}
\\
\includegraphics[width=8.5cm, height=8cm]
{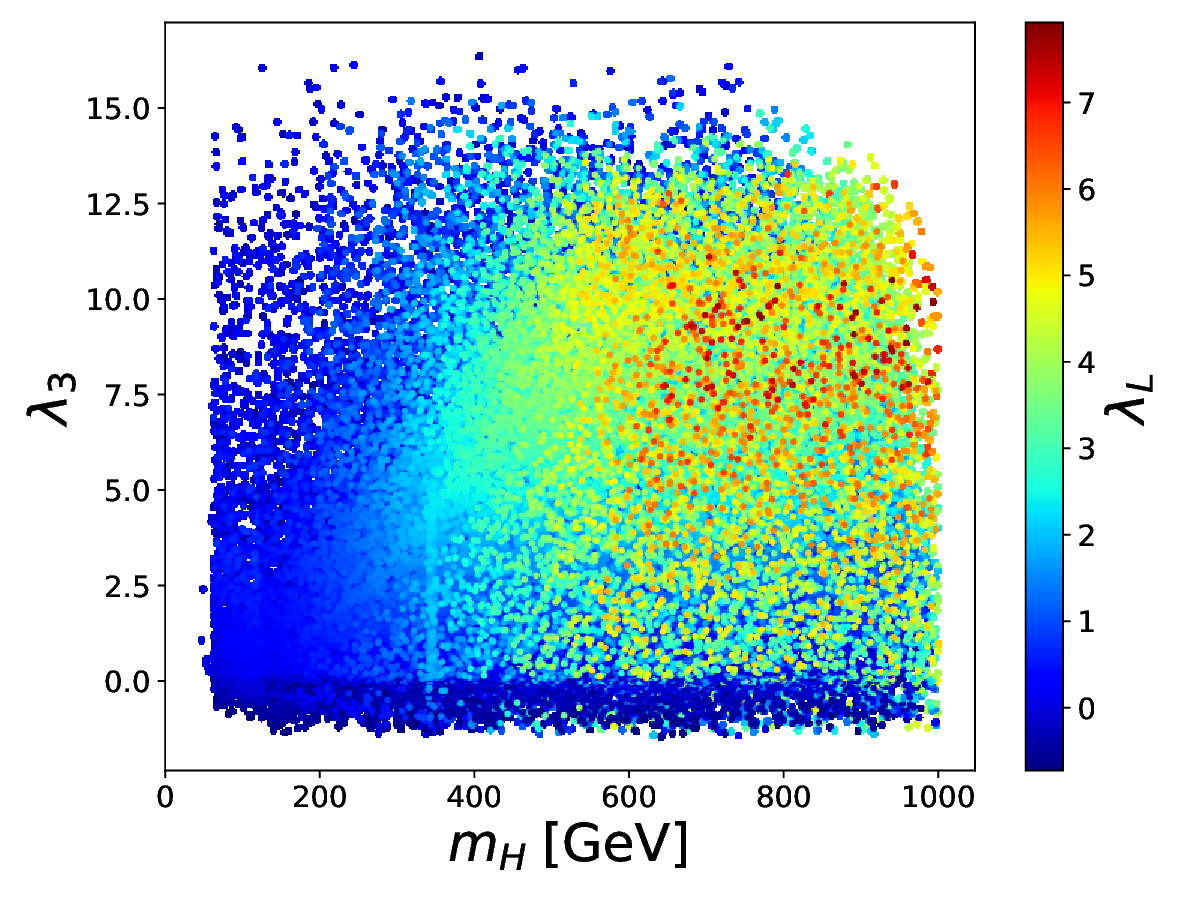}
&
\includegraphics[width=8.5cm, height=8cm]
{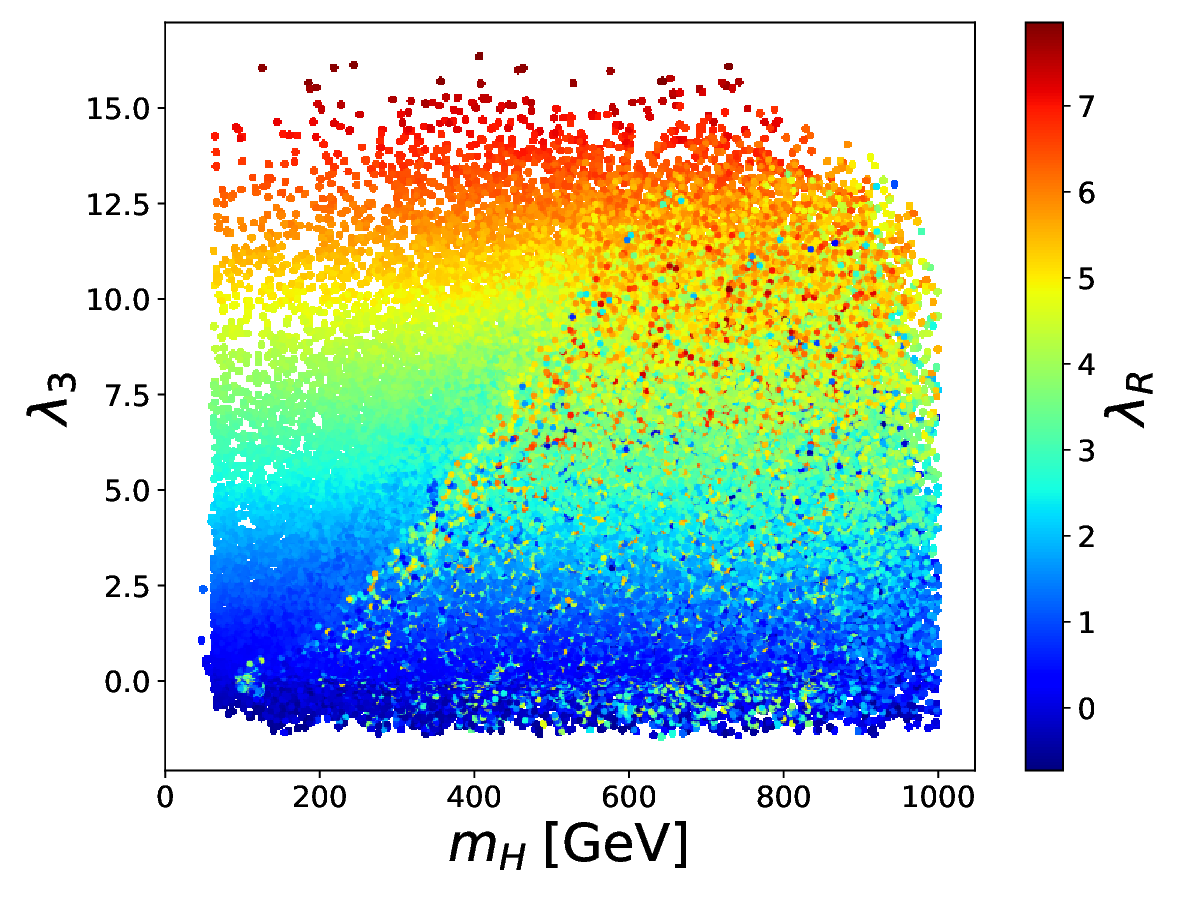}
\end{tabular}
\caption{\label{scan1}
The upper-left panel depicts the
correlation among the charged Higgs
mass $m_{H^{\pm}}$, the DM
mass $m_H$, and the scalar coupling
$\lambda_3$. In the upper-right panel,
the allowed parameter space is shown
in terms of $m_{H^{\pm}}$, $m_H$,
and the CP-odd Higgs mass $m_A$.
The lower-left panel illustrates the
allowed regions in the
$(\lambda_3,\,m_H,\,\lambda_L)$
parameter space, whereas the
lower-right panel presents the
corresponding distribution in the
$(\lambda_3,\,m_H,\,\lambda_R)$
plane.}
\end{figure}

We now outline the procedure adopted
to implement the various constraints
as presented in above
in our numerical analysis.
First, the IDM parameter space is
randomly scanned over the following
ranges:
\begin{eqnarray}
40~\mathrm{GeV} \leq m_H 
\leq 1000~\mathrm{GeV}, \quad
80~\mathrm{GeV} \leq m_{A/H^{\pm}}
\leq 1000~\mathrm{GeV},
\\
0 \leq \lambda_2 \leq 8\pi, \quad
-8\pi \leq \lambda_3 \leq 8\pi .
\end{eqnarray}
The generated parameter points are
first subjected to the theoretical
constraints discussed above.
Subsequently, the surviving samples
are tested against EWPOs
using {\tt 2HDMC}~\cite{Eriksson:2009ws}.
In the next step, the resulting
parameter space is further constrained
by Higgs precision measurements at
the LHC. For this purpose, we employ
the public packages {\tt HiggsSignals} 
and {\tt HiggsBounds}~\cite{Eriksson:2009ws,
Bahl:2022igd}.
Finally, all surviving parameter points are
required to satisfy the constraints from
Higgs self-coupling measurements implemented
in {\tt anyH3}, as well as the upper limits
on invisible Higgs decay branching ratios.
Having obtained the constrained
parameter space, we now investigate
the allowed regions of the model
under consideration by presenting
the corresponding scatter plots.
The scatter plots illustrate the
correlations among the relevant model
parameters within the allowed parameter
space in Fig.~\ref{scan1}. The upper-left 
panel displays
the correlation between the charged
Higgs mass $m_{H^{\pm}}$, the dark
matter mass $m_H$, and the coupling
$\lambda_3$. The upper-right panel
shows the corresponding correlation
among $m_{H^{\pm}}$, DM mass $m_H$, 
and the CP-odd Higgs mass
$m_A$.
In the lower-left panel, the allowed
regions in the parameter space of
$\lambda_3$, $m_H$, and $\lambda_L$
are presented, while the lower-right
panel shows the corresponding scatter
plot for $\lambda_3$, $m_H$, and
$\lambda_R$.
In the upper-left panel of Fig.~\ref{scan1},
it is interesting to observe that the region
with small values of $\lambda_3$ generally
corresponds to $m_H \geq m_{H^\pm}$, whereas
the region with large values of $\lambda_3$
is associated with $m_H \leq m_{H^\pm}$. The
latter scenario is particularly relevant for
our analysis, since it leads to an enhanced
$hH^{\pm}H^{\mp}$ coupling while simultaneously
allowing the inert scalar $H$
to play the role of DM candidate. 
In the upper-right panel of Fig.~\ref{scan1},
the correlations among the three inert scalar
masses are presented. It can be observed that
two of the inert scalar masses tend to be nearly
degenerate, while the remaining one can span a
wide range of values. This behavior mainly
originates from the constraints imposed by the EWPOs.
It is understandable since the inert scalar
particles directly couple to the gauge bosons.
As a result, the inert scalars are exchanged
in the loop contributions to the vector-boson
self-energies.

In the lower-left panel of Fig.~\ref{scan1},
we present the correlations
of DM mass with $\lambda_3$ and $\lambda_L$,
which are involved in the trilinear and quartic
Higgs couplings of the scalar sector. In particular,
we focus on the low-$m_H$ region associated with
large values of $\lambda_3$, since this scenario
provides the most promising signature for DM
searches in our analysis, as discussed in the
previous paragraph. The other allowed regions
correspond to heavier DM masses together with
sizable values of both $\lambda_3$ and
$\lambda_L$. In addition, for $\lambda_3 \leq 0$
and $\lambda_L \leq 0$, the scanned
DM mass can span a wide range of values.
Finally, we investigate the correlations 
of DM mass with $\lambda_3$ 
and $\lambda_R$
in the lower-right panel of Fig.~\ref{scan1}. 
We find that no strong 
correlation among these parameters can be
identified. The only noticeable feature
is that $\lambda_3$ tends to be proportional 
to $\lambda_R$, whereas DM mass can still
span a wide range of values.

The allowed parameter regions obtained
from the above theoretical and
experimental constraints are further
subjected to bounds arising from the
DM relic abundance and direct
detection experiments. The theoretical
predictions for the relic density and
the spin-independent direct-detection
cross sections are evaluated using the
public package
{\tt micrOMEGAs\_5.0.4}
~\cite{
Banerjee:2021oxc,
Belanger:2001fz,
Belanger:2004yn,
Belanger:2006is,
Belanger:2013oya,
Belanger:2018ccd}.
To avoid overclosing the Universe,
the following condition on the dark
matter relic abundance is imposed:
\begin{eqnarray}
 \Omega_H h^2 \leq \Omega_c h^2
 = 0.1200 \pm 0.0012 .
\end{eqnarray}
Here, $\Omega_c h^2$ represents the dark matter
relic abundance reported by the Planck
Collaboration~\cite{Planck:2018nkj,Planck:2018vyg}.
However, as demonstrated in
Ref.~\cite{Banerjee:2019luv},
radiative corrections may induce
non-negligible effects on the relic
density prediction. To account for
these higher-order uncertainties, we
include an additional theoretical
uncertainty of $20\%$, which is added
linearly to the experimental uncertainty
reported by PLANCK~\cite{Planck:2018vyg}.
Consequently, the relic abundance is
required to lie within the following
range:
\begin{eqnarray}
0.095 \lesssim \Omega h^2 
\lesssim 0.143.
\end{eqnarray}
Furthermore, DM direct
detection cross sections are constrained
by the latest constraints from
the LUX-ZEPLIN (LZ) experiment~\cite{LZ:2022lsv}.
In the case of a multi-component
DM scenario, we impose
the following condition:
\begin{eqnarray}
\sigma_{DD}\frac{\Omega_H}{\Omega_c}
\leq \sigma^{\rm LZ}_{\rm lim}(m_H).
\end{eqnarray}
After imposing all the constraints
discussed above, the surviving viable
parameter points in the IDM are listed
in Table~\ref{tab:idm_benchmark_points_DD}.
The first column labels the surviving
benchmark points, for which a total
of 23 benchmark scenarios are obtained.
The subsequent columns present the
corresponding parameter values of
the IDM. The last column shows the
predicted DM direct detection
cross section associated with each
benchmark point. 
Among all viable benchmark scenarios,
the surviving benchmark points from
BP3 to BP8 are particularly suitable
for our investigation of DM
signatures through charged Higgs pair
production at future muon colliders.
These benchmark points will therefore
be employed in the subsequent
phenomenological analysis.
\begin{table}[H]
\centering
\renewcommand{\arraystretch}{1.5}
\resizebox{\textwidth}{!}{%
\begin{tabular}{lccccccccc}
\hline
\hline
			PB
			& $m_H$ [GeV] 
			& $m_A$ [GeV] 
			& $m_{H^\pm}$ [GeV] 
			& $\mu_2^2$ [$\mathrm{GeV}^2$] 
			& $\lambda_2$ 
			& $\lambda_3$ 
			& $\lambda_L$ 
			& $\lambda_R$
			& $\sigma_{\rm DD}$ [pb] \\
			\hline
			\hline
			BP1  & 73.2022 & 119.125 & 103.999 & 5315.17 & 6.94452 & 0.181465 & 0.00071574 & 0.146402 & $3.3304\times10^{-12}$ \\
			BP2  & 72.0619 & 108.023 & 131.685 & 5130.79 & 5.53691 & 0.402816 & 0.0010249 & 0.107848 & $7.0439\times10^{-12}$ \\
			BP3  & 71.2797 & 105.388 & 152.768 & 5033.5 & 1.41676 & 0.603868 & 0.00077999 & 0.100175 & $4.1685\times10^{-12}$ \\
			BP4  & 71.203 & 224.382 & 249.572 & 5169.02 & 7.32215 & 1.88429 & -0.00163542 & 0.74521 & $1.8365\times10^{-11}$ \\
			BP5  & 71.6266 & 297.107 & 304.627 & 5128.14 & 2.34321 & 2.89221 & $3.677\times10^{-5}$ & 1.37146 & $9.1755\times10^{-15}$ \\
			BP6  & 71.8829 & 562.734 & 565.205 & 5231.97 & 1.88895 & 10.3662 & -0.00106919 & 5.13714 & $7.7035\times10^{-12}$ \\
			BP7  & 71.1158 & 688.885 & 695.424 & 4957.11 & 0.761784 & 15.7909 & 0.00165525 & 7.74612 & $1.8858\times10^{-11}$ \\
			BP8  & 56.6849 & 626.698 & 630.727 & 3272.39 & 1.61427 & 13.016 & -0.00097676 & 6.42442 & $1.0268\times10^{-11}$ \\
			BP9  & 681.613 & 690.137 & 682.545 & 466553 & 2.1321 & -0.0225814 & -0.0322709 & 0.160594 & $1.7991\times10^{-11}$ \\
			BP10 & 565.099 & 569.04 & 566.688 & 320897 & 0.935116 & 0.00784978 & -0.0257318 & 0.0479934 & $7.3869\times10^{-11}$ \\
			BP11 & 671.153 & 672.62 & 674.785 & 449579 & 2.74042 & 0.189891 & 0.0142999 & 0.0468188 & $1.6182\times10^{-11}$ \\
			BP12 & 588.129 & 591.549 & 591.597 & 344204 & 5.97698 & 0.190784 & 0.0279124 & 0.0944558 & $8.0257\times10^{-11}$ \\
			BP13 & 945.048 & 956.394 & 946.062 & 897259 & 2.93118 & -0.0733822 & -0.068327 & 0.287524 & $1.8649\times10^{-10}$ \\
			BP14 & 726.33 & 717.909 & 718.107 & 512536 & 1.85355 & 0.103651 & 0.247742 & 0.047137 & $1.5791\times10^{-10}$ \\
			BP15 & 800.767 & 809.193 & 804.777 & 642346 & 2.83939 & 0.175506 & -0.0184321 & 0.205325 & $1.8896\times10^{-11}$ \\
			BP16 & 946.259 & 956.112 & 951.154 & 896953 & 7.3365 & 0.255383 & -0.0255048 & 0.283681 & $2.5919\times10^{-11}$ \\
			BP17 & 944.468 & 953.104 & 948.794 & 890187 & 7.74247 & 0.330687 & 0.0302339 & 0.300529 & $3.6560\times10^{-11}$ \\
			BP18 & 823.58 & 832.386 & 828.799 & 678507 & 6.14542 & 0.27715 & -0.00367038 & 0.236855 & $7.0837\times10^{-13}$ \\
			BP19 & 908.237 & 912.256 & 915.371 & 826331 & 6.8219 & 0.381811 & -0.0236891 & 0.0969886 & $2.4269\times10^{-11}$ \\
			BP20 & 981.293 & 978.122 & 986.499 & 959260 & 2.59453 & 0.459237 & 0.0606503 & -0.0418537 & $6.4252\times10^{-11}$ \\
			BP21 & 868.164 & 876.088 & 875.146 & 754820 & 5.65023 & 0.364903 & -0.0183226 & 0.209655 & $1.5888\times10^{-11}$ \\
			BP22 & 940.048 & 941.797 & 948.44 & 885500 & 6.36023 & 0.463099 & -0.0298723 & 0.0244257 & $3.6026\times10^{-11}$ \\
			BP23 & 986.261 & 989.105 & 995.51 & 973395 & 1.07937 & 0.582107 & -0.0112867 & 0.0813799 & $4.6728\times10^{-12}$ \\
\hline
\hline
\end{tabular}
}
\caption{\label{tab:idm_benchmark_points_DD} 
The benchmark points of the IDM that
survive all theoretical, experimental,
and DM constraints are
presented in this Table.
We emphasize that the surviving benchmark
points from BP3 to BP8 are particularly
well suited for our investigation of
DM signatures arising from
charged Higgs pair production at future
muon colliders.}
\end{table}
\section{Charged Higgs pair 
production at 
future multi-TeV muon colliders
}
In this section, we investigate charged
Higgs boson pair production at future
multi-TeV muon colliders, focusing on
the following signal channels:
$
\mu^- \mu^+ \to
\nu_{\mu} \bar{\nu}_{\mu}
H^{\pm} H^{\mp}
\to
\ell^+ \ell^- + \rm{missing\; energy},
$
$
\mu^- \mu^+ \to
\nu_{\mu} \bar{\nu}_{\mu}
H^{\pm} H^{\mp}
\to
\ell^{\pm} + 2\,\rm{jets}
+ \rm{missing\; energy},
$
and
$
\mu^- \mu^+ \to
\nu_{\mu} \bar{\nu}_{\mu}
H^{\pm} H^{\mp}
\to
4\,\rm{jets}
+ \rm{missing\; energy}
$ with $\ell =e, \mu$. 
Within the scope of this work, the
lightest inert scalar particle $H$
is assumed to constitute the DM 
candidate. The numerical
computations are performed 
using {\tt MadGraph5\_aMC@NLO}~\cite{Alwall:2014hca}. 
In the following computations, 
all SM input parameters are taken 
from the Particle Data Group~\cite{PDG}.
\subsection{$\mu^- \mu^+ \to
\nu_{\mu} \bar{\nu}_{\mu} 
H^{\pm} H^{\mp}
\to 
4\; \rm {jets}
+ \rm{mising\; energy}$}
We first consider the most relevant
signal channel,
$
\mu^- \mu^+ \to
\nu_{\mu} \bar{\nu}_{\mu}
H^{\pm} H^{\mp}
\to
4\,\mathrm{jets}$ plus
mising energy.
The final state is characterized by four jets accompanied by missing energy. In this production channel, we already considered
the decay mode $H^{\pm}\to (W^{\pm}\to 2\mathrm{jets})H$
which gives rise to a final state containing four jets and missing energy originating from DM ($H$) and neutrinos.
For the signal events under consideration, the corresponding SM background processes included in the analysis are listed in Table~\ref{SM4jets}. These backgrounds can be classified into two categories. The first category comprises processes involving gluon-induced dijet production accompanied by additional electroweak interactions. The second category consists of purely electroweak processes, in which vector bosons and the SM-like Higgs boson decay into dijet final states.

The last two background channels listed 
in Table~\ref{SM4jets}, which involve 
the decay of the SM-like Higgs boson into 
a pair of bottom quarks, are expected to yield 
subleading contributions relative to the other background processes considered in this study.
\begin{table}[H]
\centering
\begin{tabular}{lll}
\hline\hline
Category & Process & Description \\
\hline\hline
QCD+EW 
& $\mu^+ \mu^- \to \nu_\ell 
\bar{\nu}_\ell j j j j$ 
& Includes $g\to 2\textrm{jets}$
contributions \\

\hline
Pure EW 
& $\mu^+ \mu^- \to \nu_\ell
\bar{\nu}_\ell W^+ W^-$ 
& $W^\pm \to j j$ \\

& $\mu^+ \mu^- \to \nu_\ell \bar{\nu}_\ell Z Z$ 
& $Z \to j j$ \\

& $\mu^+ \mu^- \to \nu_\ell \bar{\nu}_\ell hh$ 
& $h \to b\bar{b}$ (treated as jets) \\

& $\mu^+ \mu^- \to \nu_\ell \bar{\nu}_\ell Z h$ 
& Mixed $Z/h \to j j$ channels \\
\hline\hline
\end{tabular}
\caption{
\label{SM4jets}
Standard Model background processes
contributing to
$
\mu^+ \mu^- \to
\nu_\ell \bar{\nu}_\ell
+ 4\,\text{jets}.
$
The two processes
$
\mu^+ \mu^- \to
\nu_\ell \bar{\nu}_\ell hh
$
and
$
\mu^+ \mu^- \to
\nu_\ell \bar{\nu}_\ell Zh
$
are expected to provide smaller
contributions compared with the
remaining background channels.
}
\end{table}
In this analysis, a cut-based event
selection is applied to both the
signal and SM background samples
according to the following criteria:
\begin{eqnarray}
p_T(j_i) \geq 20~\mathrm{GeV}, 
\quad |\eta(j_i)| \leq 5.0,
\quad \Delta R(j_i,j_j) \geq 0.4,
\quad i,j = 1, \ldots, 4.
\end{eqnarray}
After imposing the above cut-based
selection criteria, the differential
cross sections with respect to the
relevant kinematic variables defined
in Appendix~B are presented in
Fig.~\ref{cutbases4jets}. These
differential distributions are
subsequently employed as input
features for the ML
training procedure, which significantly
improves the sensitivity to the signal
processes.
In the presented figures, the light-blue
histograms correspond to the SM
background, whereas the orange
histograms represent the signal
contribution. We observe that the
signal and SM background can be
efficiently discriminated in several
of these kinematic distributions,
thereby demonstrating the strong
discriminating power of the selected
observables. 

The significance for the events 
is calculated by~\cite{Braathen:2024ckk}
\begin{eqnarray}
\label{ZZ}
 Z= \sqrt{2(S+B)\log(1+\dfrac{S}{B})-S}
\end{eqnarray}
where $S$ ($B$) denotes the signal
(background) events. 
The expected event yields for the signal and background processes are computed as
$\mathcal{L}\times \sigma_{S/B}$,
where $\mathcal{L}$ denotes the integrated luminosity and $\sigma_{S/B}$ represents the corresponding signal/background cross section.
\begin{figure}[H]
\centering
\begin{tabular}{c}
\includegraphics[width=16cm, height=20cm]
{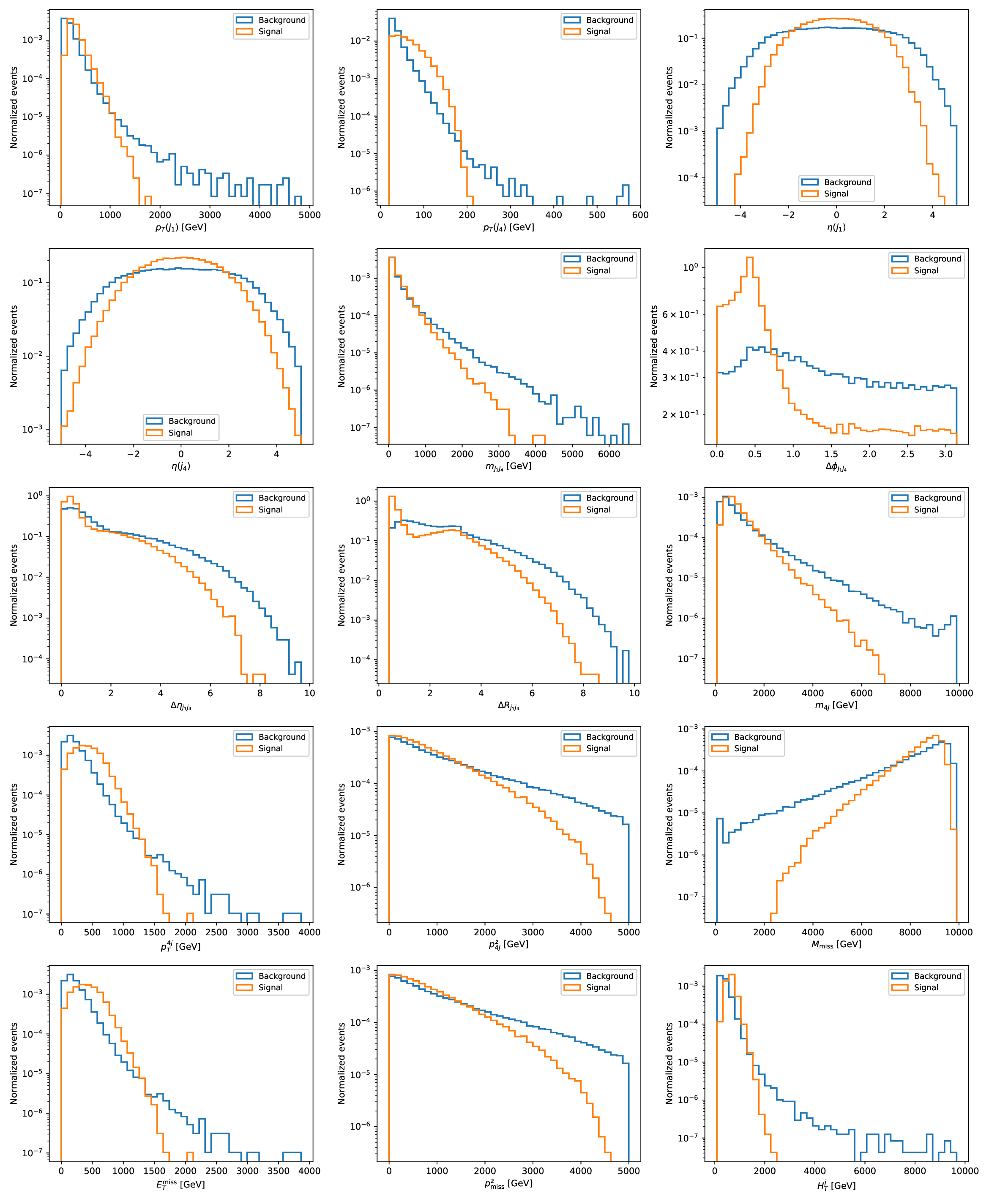}
\end{tabular}
\caption{\label{cutbases4jets} 
Distributions of the most relevant
kinematic observables after imposing
the selection cut-based on both the signal
and SM background processes. These
datasets are subsequently employed
as input samples for the machine-
learning training procedure.
}
\end{figure}
We first present the signal significance
and the corresponding SM background
yields after applying the basic
selection cuts, as well as following
the ML analysis. The
results are generated at a center-of-mass
energy of $\sqrt{s}=3~\mathrm{TeV}$
with an integrated luminosity of
$\mathcal{L}=1000~\mathrm{fb}^{-1}$, as 
shown in Table~\ref{4jets}.
Our results indicate that the signal significances remain below unity for all benchmark points BP4--BP8, primarily due to the large  SM background contributions. In contrast, BP3 yields a statistical significance exceeding the $5\sigma$ discovery threshold, thereby offering promising discovery potential at the considered collider setup.
\begin{table}[H]
\centering
\begin{tabular}
{l@{\hspace{1.5cm}}l@{\hspace{1.5cm}}l
@{\hspace{1.5cm}}l@{\hspace{1.5cm}}l
@{\hspace{1.5cm}}l}
\hline
\hline
BP
& $S_{\rm base}$ 
& $B_{\rm base}$ 
& $S_{\rm ML}$ 
& $B_{\rm ML}$ 
& $Z$ \\
\hline\hline
BP3 &  240.3   & 45010    & 102.9   & 180.040   & 7.073  \\
BP4 &  24.20   & 45010    & 12.32   & 3393.75 & 0.211    \\
BP5 &  26.20   & 45010    & 13.52   & 2822.13 & 0.254    \\
BP6 &  22.28   & 45010    & 9.331    & 652.645  & 0.364  \\
BP7 &  10.13   & 45010    & 2.969    & 184.541  & 0.218  \\
BP8 &  16.18   & 45010    & 4.571    & 220.549  & 0.307  \\
\hline
\hline
\end{tabular}
\caption{\label{4jets}
The signal
significance and the corresponding SM
background after applying the basic
selection cuts, as well as after the
machine learning analysis. 
The data are generated at $\sqrt{s}=3$ TeV
with an integrated luminosity of
$\mathcal{L}=1000~\mathrm{fb}^{-1}$.
}
\end{table}

In Table~\ref{4jets10teV}, we present the
number of signal events versus the
SM background at $\sqrt{s}=10$ TeV
with the integrated luminosity $\mathcal{L}=10,000$fb$^{-1}$. The first column lists
the benchmark points selected in
Table~\ref{tab:idm_benchmark_points_DD}.
The second and third columns show the
signal and SM background after applying
the basic selection cuts. The fourth
and fifth columns correspond to the
signal and SM background after the
ML training and
the application of optimized cuts 
from ML analysis.
The last column presents the signal
significance, calculated using
Eq.~\ref{ZZ} with applying cuts 
from the ML analysis. 
With the help of the ML strategies,
we find that the statistical 
significance $Z$ exceeds $5\sigma$ 
in almost all cases, 
except for BP4 and BP5. 
\begin{table}[H]
\centering
\begin{tabular}{l@{\hspace{1.5cm}}
l@{\hspace{1.5cm}}
l@{\hspace{1.5cm}}
l@{\hspace{1.5cm}}
l@{\hspace{1.5cm}}l}
\hline
\hline
BP
& $S_{\rm base}$ 
& $B_{\rm base}$ 
& $S_{\rm ML}$ 
& $B_{\rm ML}$ 
& $Z$ \\
\hline\hline
BP3 &  6390  & 1338000    & 1579.61  & 1070.40  & 40.564   \\
BP4 &  814.2 & 1338000    & 414.265  & 87237.6  & 1.4010   \\
BP5 &  1063  & 1338000    & 449.968  & 51914.4  & 1.9720   \\
BP6 &  3983  & 1338000    & 1647.77  & 21274.2  & 11.156   \\
BP7 &  5527  & 1338000    & 1783.01  & 9901.20  & 17.418   \\
BP8 &  5110  & 1338000    & 2288.77  & 22478.4  & 15.017   \\
\hline
\hline
\end{tabular}
\caption{\label{4jets10teV}
The signal
significance and the corresponding SM
background after applying the basic
selection cuts, as well as after the
machine learning analysis. 
The data are generated at $\sqrt{s}=10$ TeV
with an integrated luminosity of
$\mathcal{L}=10,000~\mathrm{fb}^{-1}$.
}
\end{table}
Finally, we present the signal
significance at a center-of-mass
energy of $\sqrt{s}=14~\mathrm{TeV}$.
In this case, the integrated luminosity
is taken to be
$\mathcal{L}=20{,}000~\mathrm{fb}^{-1}$.
We also find that the statistical
significance $Z$ exceeds $5\sigma$ in
almost all benchmark points, except
for BP4 and BP5.
\begin{table}[H]
\centering
\renewcommand{\arraystretch}{1.2}
\begin{tabular}{l@{\hspace{1.5cm}}
l@{\hspace{1.5cm}}
l@{\hspace{1.5cm}}
l@{\hspace{1.5cm}}
l@{\hspace{1.5cm}}l}
\hline\hline
BP
& $S_{\rm base}$ 
& $B_{\rm base}$ 
& $S_{\rm ML}$ 
& $B_{\rm ML}$ 
& $Z$ \\
\hline\hline
BP3 & 15246   & 3322000  & 7011.6  & 12291.40 & 58.33  \\
BP4 & 2002    & 3322000  & 54.454  & 11543.95 & 0.506  \\
BP5 & 2670    & 3322000  & 1426.9  & 201313.2 & 3.176  \\
BP6 & 11622   & 3322000  & 4253.7  & 35545.40 & 22.13  \\
BP7 & 18292   & 3322000  & 3315.6  & 6976.20  & 37.05  \\
BP8 & 15844   & 3322000  & 5328.3  & 26243.80 & 31.86  \\
\hline
\hline
\end{tabular}
\caption{\label{MLSBZ14TeV}
The signal
significance and the corresponding SM
background after applying the basic
selection cuts, as well as after the
machine learning analysis. 
The data are generated at $\sqrt{s} 
= 14$ TeV
with an integrated luminosity of
$\mathcal{L}=20,000~\mathrm{fb}^{-1}$.
}
\end{table}
\subsection{$ \mu^- \mu^+ 
\to \nu_{\mu} \bar{\nu}_{\mu} 
H^{\pm} H^{\mp} \to  
\ell^{\pm} + 2\; \rm{jets}
+ \rm{mising\; energy}$ }
We turn our attention to the
semileptonic channel
$ \mu^- \mu^+ \to
\nu_{\mu} \bar{\nu}_{\mu}
H^{\pm} H^{\mp}
\to
\ell^{\pm} + 2\,\mathrm{jets}$
plus missing energy.
In this analysis, jets are assumed to
originate from $u, d, c, s,$ and $b$
quarks, while the charged leptons are
restricted to electrons or muons.
Due to the final-state topology
$
\ell^{\pm} + 2\,\mathrm{jets}$
associated with missing energy,
all relevant SM background 
processes are listed in
Table~\ref{tab:SM_backgrounds_leptonic}.
In this Table, the first two processes
provide the dominant contributions,
while the latter two correspond to
$2 \to 6$ processes, which are expected
to yield subdominant contributions
compared with the former cases.
\begin{table}[H]
\centering
\begin{tabular}{lll}
\hline\hline
Category & Process & Description \\
\hline\hline
Semi-leptonic 
& $\mu^+ \mu^- \to \ell^+ \nu_\ell j j$ 
& $W^+ \to \ell^+ \nu_\ell,\; W^- \to j j$ \\

& $\mu^+ \mu^- \to \ell^- \bar{\nu}_\ell j j$ 
& $W^- \to \ell^- \bar{\nu}_\ell,\; W^+ \to j j$ \\

\hline
Leptonic + jets 
& $\mu^+ \mu^- \to \nu_\ell \bar{\nu}_\ell 
\ell^+ \nu_\ell j j$ 
& Additional neutrinos from EW processes \\

& $\mu^+ \mu^- \to \nu_\ell \bar{\nu}_\ell
\ell^- \bar{\nu}_\ell j j$ 
& Multi-neutrino final states. 
\\
\hline\hline
\end{tabular}
\caption{
\label{tab:SM_backgrounds_leptonic}
The Standard Model background 
processes contributing to 
$\mu^+ \mu^- \to \ell^\pm + \nu_\ell + j j$ and 
$\mu^+ \mu^- \to \nu_\ell \bar{\nu}_\ell 
+ \ell^\pm + \nu_\ell + j j$. 
The latter processes are
included tau lepton decay 
$\tau \to \ell \nu_{\ell} \nu_{\tau}$
in final states. }
\end{table}
The signal and SM background events are
generated using
{\tt MadGraph5\_aMC@NLO}~\cite{Alwall:2014hca}
and they are evaluated by applying the following
cut-based selection criteria:
\begin{eqnarray}
 p_{T}^{\ell} \geq 10 \text{GeV}, 
 \;
 p_{T}^{j} \geq 20 \text{GeV}, 
 \; |\eta_{\ell}| \leq 2.5,
 \; |\eta_{j}| \leq 5.0,
 \; \Delta R(\ell, j)\geq 0.4, 
 \; \Delta R(j_1, j_2)\geq 0.4. 
\end{eqnarray}
Following the same procedure as in the
previous analysis, we present the
differential cross sections with
respect to the kinematic variables
defined in Appendix~B, as displayed
in Fig.~\ref{cutbases-elljet}. The
same notation for the distributions
is adopted throughout these plots.
In particular, the light-blue
histograms correspond to the SM
background, whereas the orange
histograms represent the signal
contribution.
We observe that the signal and SM
background can be efficiently
distinguished in several regions of
the considered kinematic
distributions. Our results indicate
that the signal and background events
are well separated for a number of
relevant observables. These
kinematic distributions are then
employed as input features for the
ML analysis in order
to enhance the sensitivity to the
signal discovery.
\begin{figure}[H]
\centering
\begin{tabular}{c}
\includegraphics[width=16cm, height=20cm]
{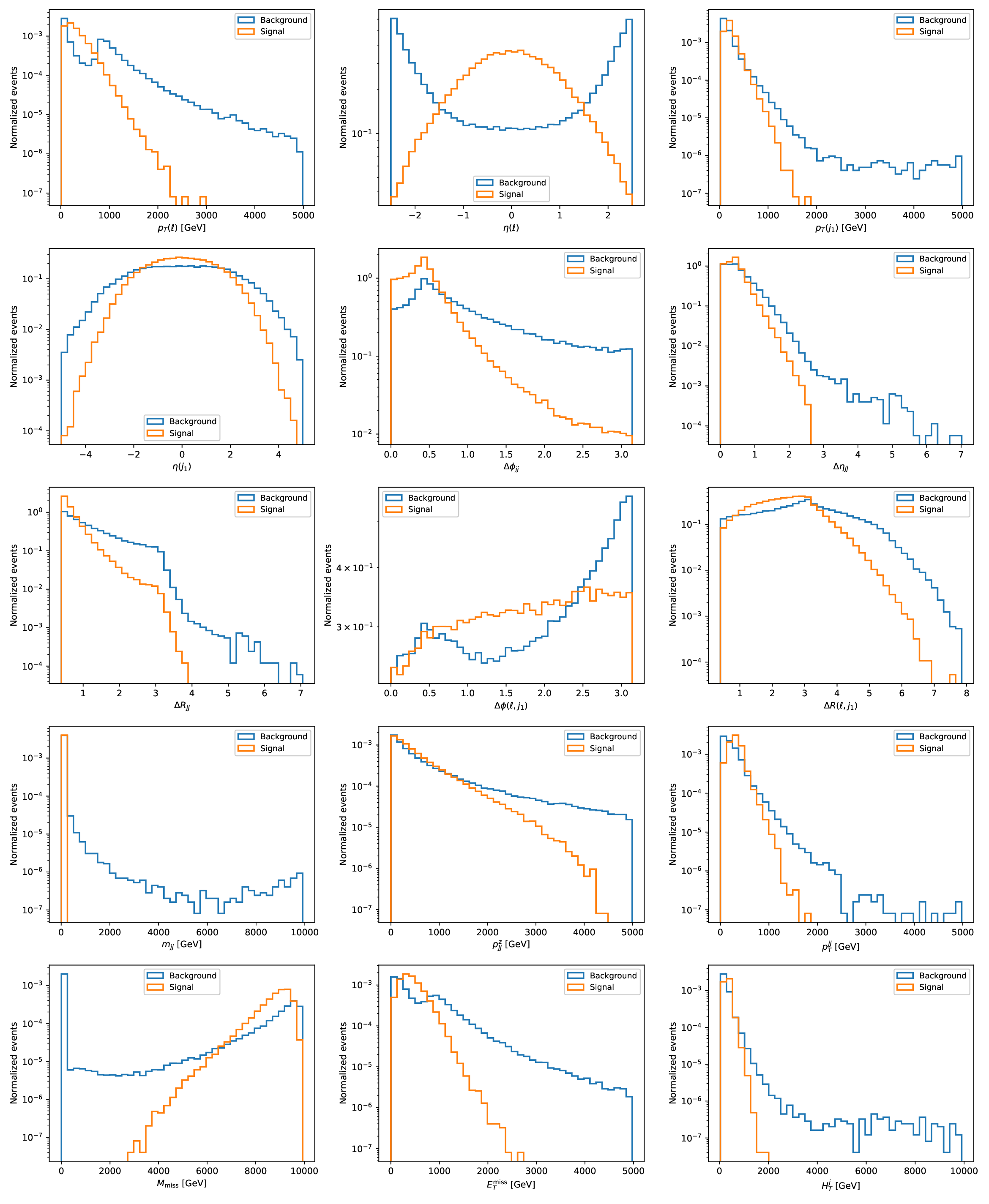}
\end{tabular}
\caption{
\label{cutbases-elljet} 
Distributions of the most relevant observables
after applying selection cuts to both signal
and background processes are presented.
These datasets are then used for machine
learning training to get the statistical
significances of the observed events at 
future multi-TeV muon collider.
}
\end{figure}
The signal significance, together with
the corresponding SM background yields,
obtained after applying the basic
selection cuts as well as the
ML analysis, is presented in
Table~\ref{ell2jet-3tev}. The event
samples are generated at
$\sqrt{s}=3~\mathrm{TeV}$ with
the integrated luminosity of
$\mathcal{L}=1000~\mathrm{fb}^{-1}$.
Our results indicate that the signal
significances for all benchmark points
from BP4 to BP8
remain below unity, mainly due to the
large SM background contributions.
In contrast, the statistical 
significance for BP3 can reach
$Z\geq 4$.
\begin{table}[H]
\centering
\begin{tabular}{l@{\hspace{1.5cm}}
l@{\hspace{1.5cm}}
l@{\hspace{1.5cm}}
l@{\hspace{1.5cm}}
l@{\hspace{1.5cm}}l}
\hline\hline
BP
& $S_{\rm base}$ 
& $B_{\rm base}$ 
& $S_{\rm ML}$ 
& $B_{\rm ML}$ 
& $Z$ \\
\hline
\hline
BP3 &  223.1  & 429600    & 82.24   & 343.680   & 4.275   \\
BP4 &  20.08  & 429600    & 11.07   & 4081.20   & 0.173   \\
BP5 &  21.19  & 429600    & 5.052   & 515.520   & 0.222   \\
BP6 &  19.71  & 429600    & 4.439   & 257.760   & 0.276   \\
BP7 &  9.938  & 429600    & 6.077   & 1632.48   & 0.150   \\
BP8 &  15.20  & 429600    & 9.512   & 1761.36   & 0.226   \\
\hline
\hline
\end{tabular}
\caption{\label{ell2jet-3tev}
The signal
significance and the corresponding SM
background after applying the basic
selection cuts, as well as after the
machine learning analysis. 
The data are generated at $\sqrt{s}=3$ TeV
with the integrated luminosity of
$\mathcal{L}=1000~\mathrm{fb}^{-1}$.
}
\end{table}
In Table~\ref{ell2jet-10tev}, we
present the signal significance and
the corresponding SM background
after applying the basic
selection cuts, as well as following
the ML analysis
at $\sqrt{s}=10$ TeV with 
$\mathcal{L}=10,000$ fb$^{-1}$.
Owing to the strong discriminating
power of the ML approach, the signal
significance exceeds $5\sigma$ for
many benchmark points listed 
in the Table, except for
BP4 and BP5.
\begin{table}[htbp]
\centering
\begin{tabular}{l@{\hspace{1.5cm}}l@{\hspace{1.5cm}}l@{\hspace{1.5cm}}l@{\hspace{1.5cm}}l@{\hspace{1.5cm}}l}
\hline
\hline
BP
& $S_{\rm base}$ 
& $B_{\rm base}$ 
& $S_{\rm ML}$ 
& $B_{\rm ML}$ 
& $Z$ \\
\hline\hline
BP3 &  5412  & 1177000    & 1672.31   & 2354    & 31.26  \\
BP4 &  636.7 & 1177000    & 301.159   & 55319   & 1.279   \\
BP5 &  824.4 & 1177000    & 529.265   & 97691   & 1.692   \\
BP6 &  3929  & 1177000    & 2518.49   & 57673   & 10.41  \\
BP7 &  8829  & 1177000    & 1509.76   & 1177    & 37.62  \\
BP8 &  6625  & 1177000    & 2530.75   & 17655   & 18.62  \\
\hline
\hline
\end{tabular}
\caption{\label{ell2jet-10tev}
The signal
significance and the corresponding SM
background after applying the basic
selection cuts, as well as after the
machine learning analysis. 
The data are generated at $\sqrt{s}=10$ TeV
with the integrated luminosity of
$\mathcal{L}=10\,000~\mathrm{fb}^{-1}$.
}
\end{table}

Table~\ref{ell2jet-14tev} summarizes
the signal significance together with
the corresponding SM background
yields after imposing the basic
selection cuts and performing the
ML analysis. The event
samples are generated at
$\sqrt{s}=14~\mathrm{TeV}$ with the
integrated luminosity of
$\mathcal{L}=20{,}000~\mathrm{fb}^{-1}$.
Our results show that the
ML framework
substantially improves the discovery
potential of the signal. In
particular, the statistical
significance exceeds the
$5\sigma$ discovery threshold for
most benchmark points presented in 
the Table, with the
exception of BP4 and BP5.
\begin{table}[H]
\centering
\begin{tabular}{l@{\hspace{1.5cm}}
l@{\hspace{1.5cm}}
l@{\hspace{1.5cm}}
l@{\hspace{1.5cm}}
l@{\hspace{1.5cm}}l}
\hline\hline
BP
& $S_{\rm base}$ 
& $B_{\rm base}$ 
& $S_{\rm ML}$ 
& $B_{\rm ML}$ 
& $Z$ \\
\hline\hline
BP3 & 12536   & 1878920  & 4644.59   & 14467.684  & 36.78   \\
BP4 & 1525.2  & 1878920  & 968.044   & 271316.05  & 1.857    \\
BP5 & 2016    & 1878920  & 1105.37   & 191649.84  & 2.522    \\
BP6 & 11476   & 1878920  & 5282.40   & 71211.068  & 19.56   \\
BP7 & 34160   & 1878920  & 14569.2   & 34384.236  & 73.82   \\
BP8 & 21180   & 1878920  & 8414.81   & 40020.996  & 40.71   \\
\hline
\hline
\end{tabular}
\caption{\label{ell2jet-14tev}
The signal
significance and the corresponding SM
background after applying the basic
selection cuts, as well as after the
machine learning analysis. 
The data are generated at $\sqrt{s}=14$ TeV
with the integrated luminosity of
$\mathcal{L}=20,000~\mathrm{fb}^{-1}$.
}
\end{table}
\subsection{$ \mu^- \mu^+ \to
\nu_{\mu} \bar{\nu}_{\mu}  
H^{\pm} H^{\mp} \to  
\ell^+ \ell^- + \rm{mising\; energy}$}
\begin{figure}[H]
\centering
\begin{tabular}{c}
\includegraphics[width=16cm, height=20cm]
{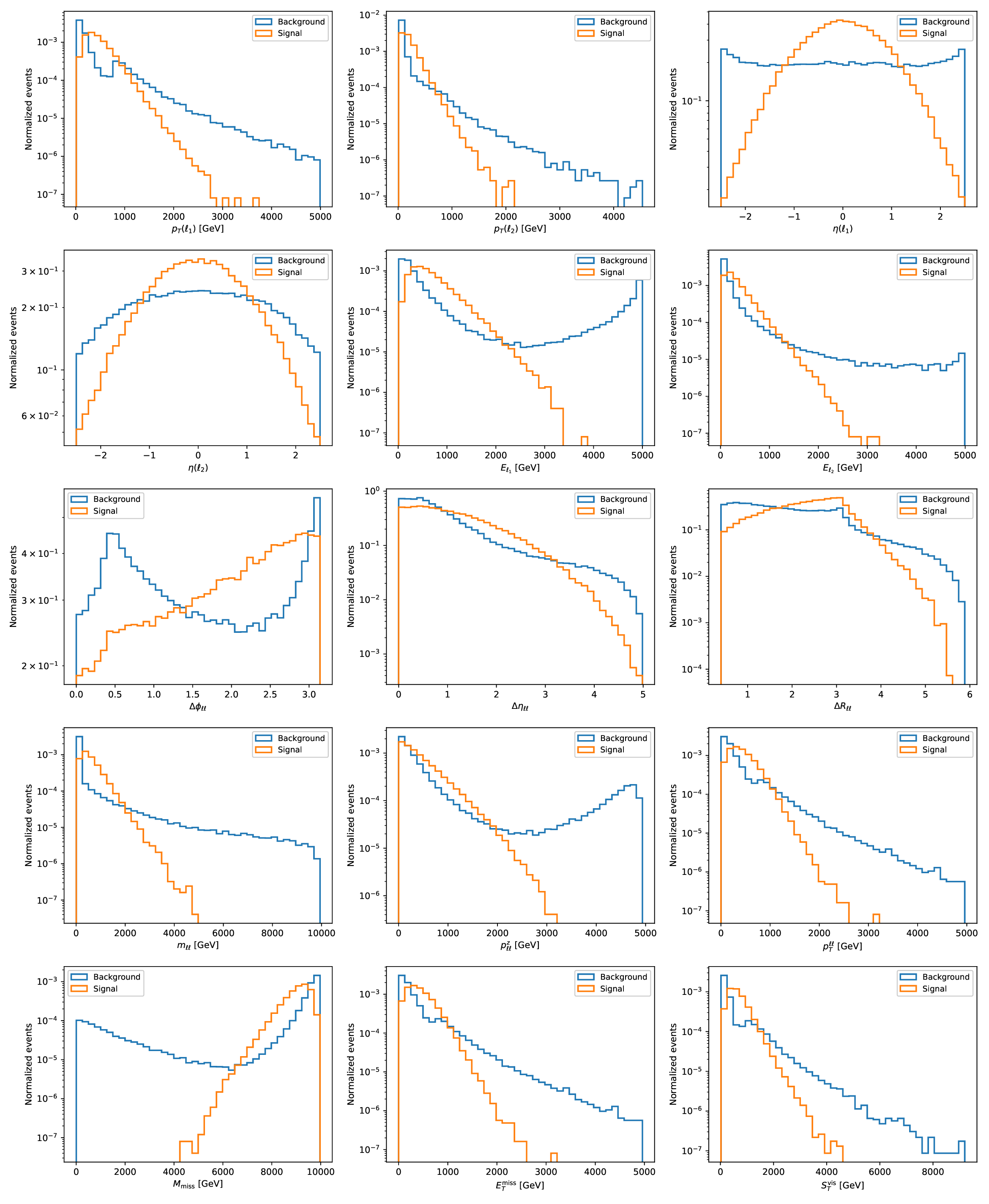}
\end{tabular}
\caption{\label{ellell-10tevdistrubution} 
Distributions of the most relevant observables
after applying selection cuts to both signal
and background processes are presented.
These datasets are then used for machine
learning training.
}
\end{figure}

We finally consider the most
challenging signal channel,
$
\mu^- \mu^+ \to
\nu_{\mu} \bar{\nu}_{\mu}
H^{\pm} H^{\mp}
\to
\ell^+ \ell^-
+ \rm{missing\; energy}.
$
Due to the final-state 
signature characterized by
$
\ell^+ \ell^-
+ \rm{missing\; energy}$,
we expect substantial SM background
contributions as summarized in
Table~\ref{SMBgllvv}.
\begin{table}[H]
\centering
\begin{tabular}{lll}
\hline\hline
Category & Process & Description \\
\hline\hline
Di-boson, VBF
& $\mu^+ \mu^- \to \ell^+ \ell^- \nu \bar{\nu}$ 
& $\ell^+ \ell^- + {\rm missing \; energy}$ \\
\hline
Tri-boson, VBF
& $\mu^+ \mu^- \to \ell^+ \ell^- \nu \bar{\nu}\nu \bar{\nu}$ 
& $\ell^+ \ell^- + {\rm missing \; energy}$ (multi-neutrino) \\

\hline\hline
\end{tabular}
\caption{
\label{SMBgllvv}
Standard Model background processes for 
$\mu^+ \mu^- \to \ell^+ \ell^- +  {\rm missing \; energy}$. 
The dominant contributions 
are from VBF processes.
}
\end{table}
Following the same procedure as in
the previous analyses, we apply the
following cut-based selection
criteria to both the signal and SM
background events:
\begin{eqnarray}
 p_{T}^{\ell} \geq 10 \text{GeV}, 
 \quad |\eta_{\ell^\pm}| \leq 2.5,
 \quad \Delta R(\ell^+, \ell^-)\geq 0.4. 
\end{eqnarray}
We then present the kinematic
distributions for both the signal
and SM background processes,
employing the same line notation
for the signal and background
contributions. The distributions are
generated at a center-of-mass energy
of $\sqrt{s}=10~\mathrm{TeV}$ with
an integrated luminosity of
$\mathcal{L}=10{,}000~\mathrm{fb}^{-1}$,
as shown in
Fig.~\ref{ellell-10tevdistrubution}.
It is also interesting to observe that
the signal and SM background can be
efficiently discriminated in several
kinematic distributions. The
differential cross sections with
respect to the considered kinematic
variables are subsequently employed
as input features for the
ML training procedure
in order to enhance the sensitivity
to the signal discovery.

We next evaluate the signal
significance at a center-of-mass
energy of $\sqrt{s}=3~\mathrm{TeV}$
with an integrated luminosity of
$\mathcal{L}=1000~\mathrm{fb}^{-1}$.
The corresponding results are
presented in
Table~\ref{ellellZ3TeV}. As expected,
owing to the large SM background
contributions, the statistical
significance remains below unity
for all considered benchmark points.

The signal significance together with
the corresponding SM background
yields, obtained after applying the
basic selection cuts as well as the
ML analysis, are
presented in
Table~\ref{ellellZ10TeV}. The event
samples are generated at a
center-of-mass energy of
$\sqrt{s}=10~\mathrm{TeV}$ with an
integrated luminosity of
$\mathcal{L}=10{,}000~\mathrm{fb}^{-1}$.
We find that the statistical
significance $Z$ exceeds the
$5\sigma$ discovery threshold for
the benchmark points BP6--BP8,
whereas the remaining benchmark
points yield significances below
the $5\sigma$ level.

\begin{table}[H]
	\centering
	\label{tab:ml-S-B-Z-two-leptons-3tev}
	\renewcommand{\arraystretch}{1.2}
	\begin{tabular}{l@{\hspace{1.5cm}}
l@{\hspace{1.5cm}}
l@{\hspace{1.5cm}}
l@{\hspace{1.5cm}}
l@{\hspace{1.5cm}}l}
\hline\hline
BP
& $S_{\rm base}$ 
& $B_{\rm base}$ 
& $S_{\rm ML}$ 
& $B_{\rm ML}$ 
& $Z$ \\
\hline
\hline
BP3 & 52.31  & 291900    & 28.765  & 4699.59   & 0.419  \\
BP4 & 4.198  & 291900    & 1.728   & 1984.92   & 0.039  \\
BP5 & 4.341  & 291900    & 1.475   & 992.460   & 0.047  \\
BP6 & 4.429  & 291900    & 0.856   & 175.140   & 0.065  \\
BP7 & 2.483  & 291900    & 0.666   & 262.710   & 0.041  \\
BP8 & 3.634  & 291900    & 1.874   & 1167.600  & 0.055  \\
\hline
\hline
\end{tabular}
\caption{\label{ellellZ3TeV}
The signal
significance and the corresponding SM
background after applying the basic
selection cuts, as well as after the
machine learning analysis. 
The data are generated at 
$\sqrt{s}=3$ TeV
with an integrated luminosity of
$\mathcal{L}=1000~\mathrm{fb}^{-1}$.
}
\end{table}

\begin{table}[H]
\centering
\begin{tabular}{l@{\hspace{1.5cm}}l@{\hspace{1.5cm}}l@{\hspace{1.5cm}}l@{\hspace{1.5cm}}l@{\hspace{1.5cm}}l}
\hline
\hline
BP
& $S_{\rm base}$ 
& $B_{\rm base}$ 
& $S_{\rm ML}$ 
& $B_{\rm ML}$ 
& $Z$ \\
\hline
BP3 &  1184  & 1795000    & 722.24    & 48465   & 3.273   \\
BP4 &  127.7 & 1795000    & 52.612    & 12565   & 0.470   \\
BP5 &  164.3 & 1795000    & 70.320    & 8975    & 0.741   \\
BP6 &  1040  & 1795000    & 609.44    & 10770   & 5.818   \\
BP7 &  4467  & 1795000    & 2328.662  & 5385    & 29.78   \\
BP8 &  2315  & 1795000    & 858.865   & 3590    & 13.81   \\ 
\hline \hline
\end{tabular}
\caption{\label{ellellZ10TeV}
The signal
significance and the corresponding SM
background after applying the basic
selection cuts, as well as after the
machine learning analysis. 
The data are generated at $\sqrt{s}=10$ TeV
with an integrated luminosity of
$\mathcal{L}=10\,000~\mathrm{fb}^{-1}$.
}
\end{table}
Table~\ref{ellellZ14TeV} summarizes
the signal significance and the
corresponding SM background yields
after imposing the basic selection
cuts and performing the
ML analysis. The event
samples are generated at a
center-of-mass energy of
$\sqrt{s}=14~\mathrm{TeV}$ with an
integrated luminosity of
$\mathcal{L}=20{,}000~\mathrm{fb}^{-1}$.
Our numerical results indicate that
the statistical significance $Z$
surpasses the $5\sigma$ discovery
threshold for the benchmark points
BP6--BP8, while the remaining
benchmark scenarios lead to
significances below the
$5\sigma$ level.
\begin{table}[H]
\centering
\begin{tabular}
{l@{\hspace{1.5cm}}l@{\hspace{1.5cm}}l
@{\hspace{1.5cm}}l@{\hspace{1.5cm}}l
@{\hspace{1.5cm}}l}
\hline
\hline
BP
& $S_{\rm base}$ 
& $B_{\rm base}$ 
& $S_{\rm ML}$ 
& $B_{\rm ML}$ 
& $Z$ \\
\hline\hline
BP3 & 2694   & 3632000   & 2209.62   & 344313.6   & 3.762  \\
BP4 & 301.2  & 3632000   & 126.805   & 74819.20   & 0.463  \\
BP5 & 398    & 3632000   & 180.453   & 63196.80   & 0.717  \\
BP6 & 3076   & 3632000   & 2154.43   & 98427.20   & 6.842  \\
BP7 & 16636  & 3632000   & 6211.88   & 14891.20   & 47.87 \\
BP8 & 7634   & 3632000   & 4810.18   & 59564.80   & 19.45 \\
\hline
\hline
\end{tabular}
\caption{\label{ellellZ14TeV}
The signal
significance and the corresponding SM
background after applying the basic
selection cuts, as well as after the
machine learning analysis. 
The data are generated at $\sqrt{s}=14$ TeV
with an integrated luminosity of
$\mathcal{L}=20,000~\mathrm{fb}^{-1}$.
}
\end{table}

\section{Conclusions}
In this work, we have investigated
DM searches through charged Higgs 
boson pair production
at future multi-TeV muon colliders
within the framework of the IDM. 
The viable parameter space of 
the IDM was first examined by 
incorporating both theoretical 
requirements and the most
recent experimental constraints.
Subsequently, the DM relic
abundance and direct-detection bounds
were evaluated within the allowed
parameter space. The surviving
parameter points satisfying all DM constraints were then employed in 
the phenomenological analysis of
charged Higgs boson pair production
followed by their decays into SM 
particles accompanied by DM 
at future multi-TeV muon colliders. 
In particular, we analyzed 
the following signal channels:
$
\mu^- \mu^+ \to
\nu_{\mu} \bar{\nu}_{\mu}
H^{\pm} H^{\mp}
\to
\ell^+ \ell^-
+
\nu_{\mu} \bar{\nu}_{\mu}
\nu_{\ell} \bar{\nu}_{\ell} HH,
$
$
\mu^- \mu^+ \to
\nu_{\mu} \bar{\nu}_{\mu}
H^{\pm} H^{\mp}
\to
\ell^\pm + 2\,\text{jets}
+ \nu_{\mu} \bar{\nu}_{\mu}
\nu_{\ell} HH,
$
and
$
\mu^- \mu^+ \to
\nu_{\mu} \bar{\nu}_{\mu}
H^{\pm} H^{\mp}
\to
4\,\text{jets}
+ \nu_{\mu} \bar{\nu}_{\mu} HH$.
The signal significance was evaluated
against the corresponding SM
backgrounds using both conventional
cut-based methods and ML
techniques. Our analysis demonstrates
that the ML framework
provides a substantial improvement in
the sensitivity to the signal channels
compared with the traditional cut-based
approach. Furthermore, the obtained
results indicate that DM
signatures arising from charged Higgs
boson pair production can be indirectly
explored with statistical significances
exceeding the $5\sigma$ discovery
threshold for several viable benchmark
points at future multi-TeV muon
colliders.
\section*{ML interpretation plots}
In this appendix, we provide an 
interpretation of the XGBoost~\cite{xgboost,
Shapley} 
classifier employed for signal--background
discrimination. As an illustrative example, 
we consider 
the following processes
$\mu^- \mu^+ \to \nu_\mu \bar{\nu}_{\mu}
H^\pm H^\mp$ 
$\to 4\; \textrm{jets}
+\nu_{\mu}\bar{\nu}_\mu HH$
for the benchmark point BP7 at
$\sqrt{s}=10~{\rm TeV}$
with the integrated luminosity of
$\mathcal{L}=10,000~{\rm fb}^{-1}$.
The corresponding final state consists 
of four visible jets accompanied by missing 
transverse energy arising from the two 
DM particles $H$ and the neutrinos.
For the ML training, all differential 
distributions shown in Fig.~\ref{cutbases4jets} for the processes under investigation are used as input features. 
To identify the kinematic observables that 
play the most significant role in the classification
procedure, we employ SHAP values~\cite{Grojean:2020ech,
Alasfar:2022vqw,Grojean:2022mef,
Bahl:2023qwk,Lundberg}. For a given 
event $x$, the SHAP value associated with a 
feature $m$ is defined as
\begin{eqnarray}
\phi_m(x) &=&
\sum_{S\subseteq F\setminus\{m\}}
\frac{|S|!\left(|F|-|S|-1\right)!}{|F|!}
\left[
v_x(S\cup\{m\})-v_x(S)
\right],
\end{eqnarray}
where $F$ denotes the full 
set of input observables, 
while $S$ represents a subset of observables that does not contain the feature $m$. The modes 
$v_x(S\cup\{m\})$ and $v_x(S)$ represent the predictions of models trained on the corresponding feature subsets.
The SHAP value $\phi_m(x)$ 
quantifies the contribution of the feature $m$ to 
the classifier output for a given event $x$, averaged over all possible feature subsets. In practice, this quantity is not computed by retraining the classifier for every possible subset. Since the classifier is based on XGBoost, the SHAP values are evaluated using the TreeExplainer algorithm implemented in the SHAP package.

The classifier output can be expressed as 
the sum of a baseline value and the SHAP 
contributions from all input features.
\begin{align}
f(x)=E[f(X)]+\sum_{m\in F}\phi_m(x).
\end{align}
In the convention adopted here, positive values of $f(x)$ correspond to signal-like events, whereas negative values correspond to background-like events. Consequently, a positive SHAP value shifts the event classification toward the signal region, while a negative SHAP value drives it toward the background region.

\begin{figure}[H]
\centering
\begin{tabular}{c}
\includegraphics[width=10cm, height=15cm]
{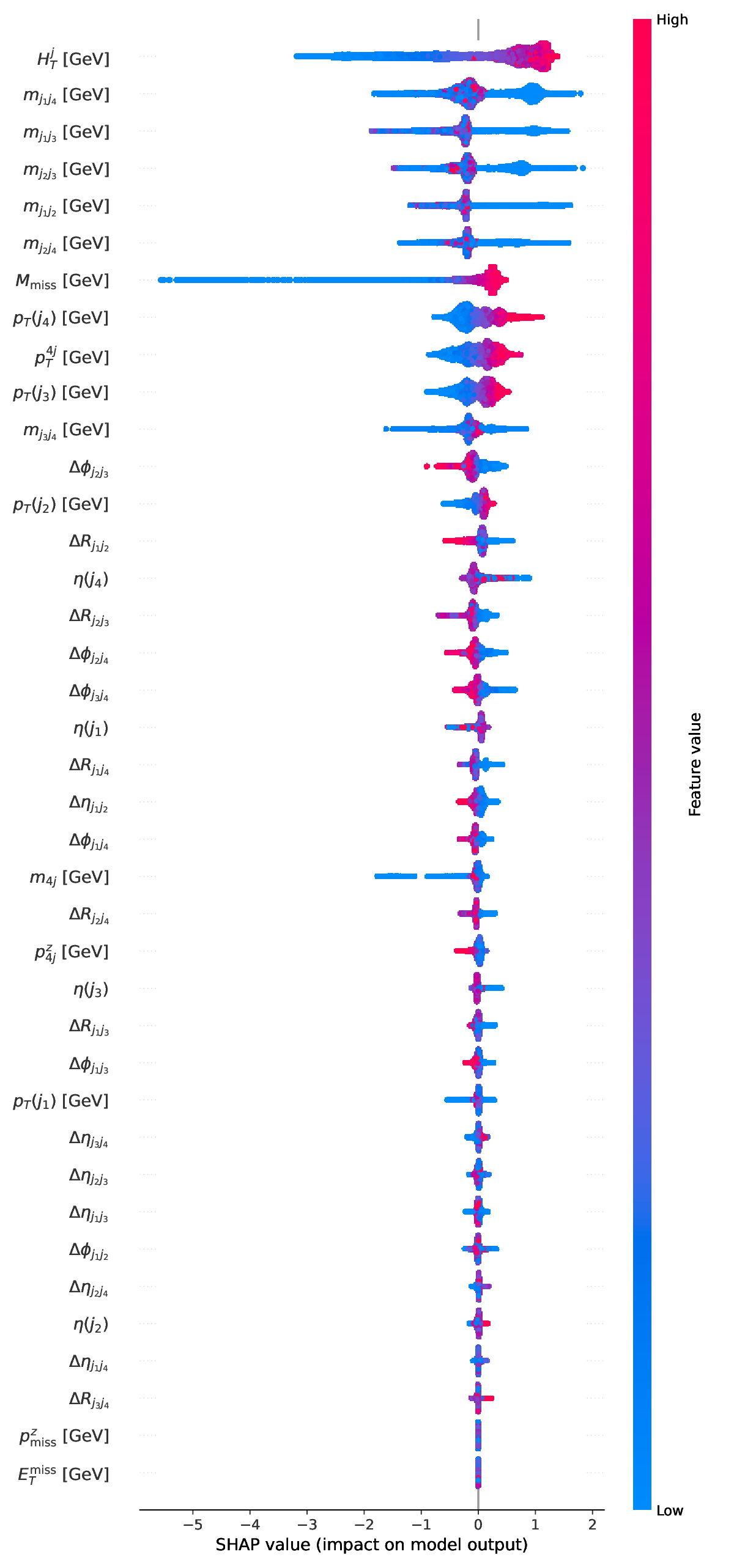}
\end{tabular}
\caption{\label{ML1}
SHAP values evaluated for all observables
considered in our ML analysis using a 
collection of events.
}
\end{figure}

Fig.~\ref{ML1} presents the SHAP summary plot for the XGBoost classifier. Each point corresponds to an event in the test sample, while its horizontal position represents the SHAP value associated with the corresponding feature. The colour presents for the numerical value of the feature. In detail, 
the red points indicate large feature values, whereas blue points correspond to small feature values. The features are ordered according to their average impact on the classifier output.
The most influential observables are the sum of the jet transverse momenta, $H_T^j$, the invariant masses of jet pairs, 
and the missing invariant mass $M_{\rm miss}=\sqrt{p_{\mathrm{miss}}
\cdot p_{\mathrm{miss}} }$
where $p_{\mathrm{miss}}$ is defined
as in Eq.~\ref{pmiss}.
In particular, $H_T^j$ exhibits the largest impact on the classifier response. Large values of $H_T^j$ tend to shift events toward the signal-like region, consistent with the harder jet activity expected from charged Higgs pair production at multi-TeV energies. The dijet invariant masses $m_{j_i j_k}$ for $i,k=1,2,3,4$ 
also play a significant role, since the signal topology contains two hadronic decay chains originating from $H^\pm\to W^\pm H$. In addition, $M_{\rm miss}$ provides substantial discriminating power because the signal contains two invisible DM particles together with neutrinos in the final state. In particular, large missing invariant masses are a characteristic feature of the signal rather than the SM background, due to the presence of DM particles in the final state.

\begin{figure}[H]
\centering
\begin{tabular}{cc}
\includegraphics[width=8cm,height=14cm]
{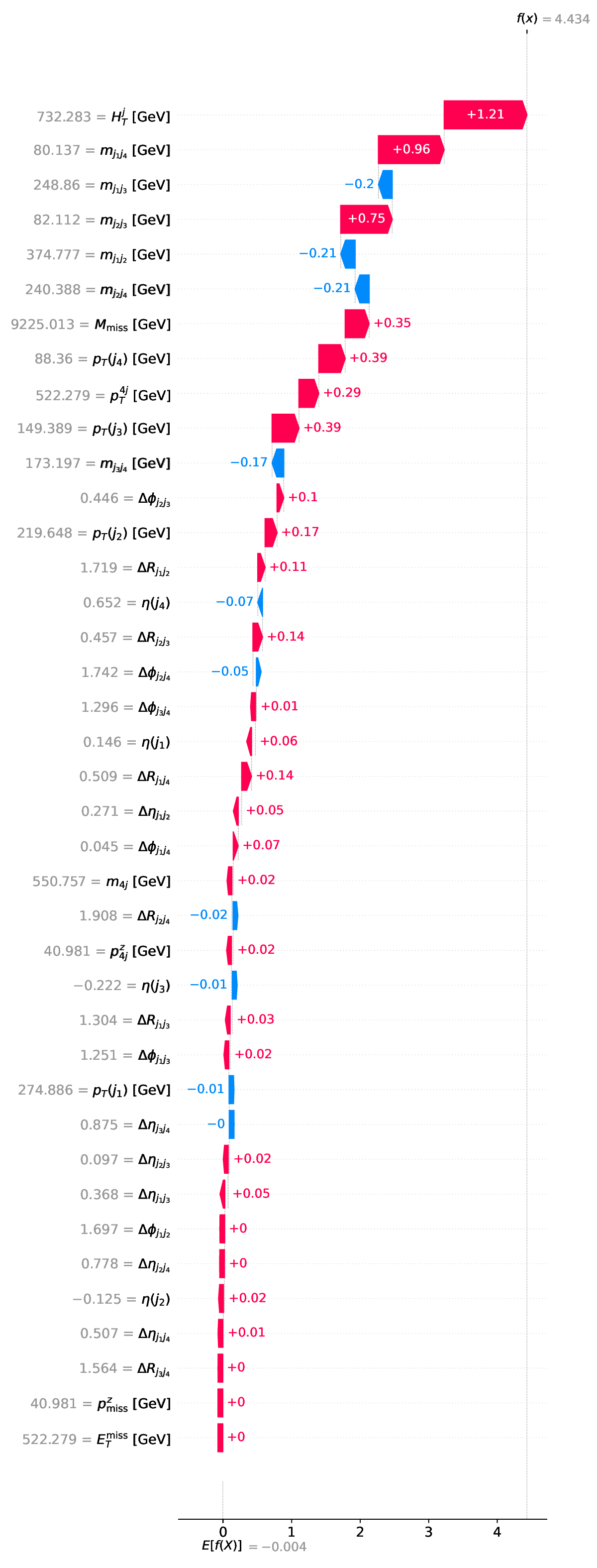}
&
\includegraphics[width = 8cm, height=14cm]
{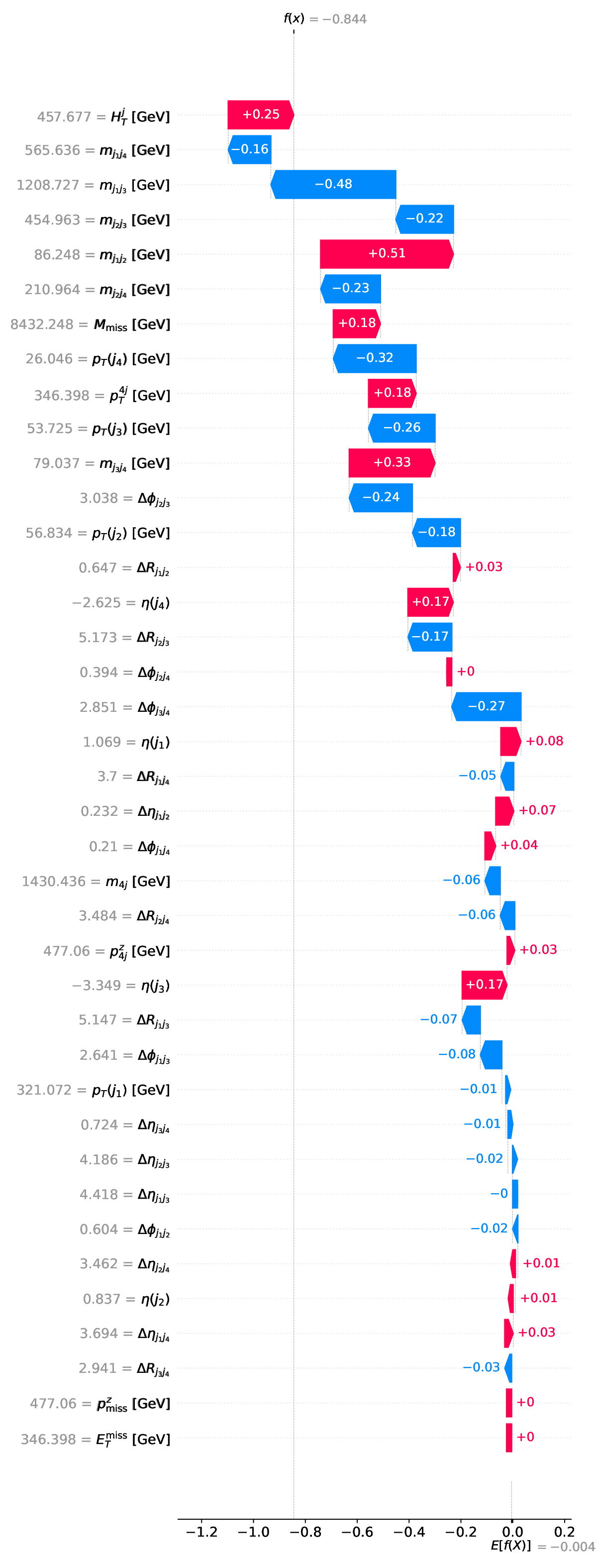}
\end{tabular}
\caption{\label{ML2}
SHAP values for a representative signal 
point (left panel) and a representative 
background point (right panel).
}
\end{figure}

Fig.~\ref{ML2} shows the SHAP waterfall plots 
for one signal-like event and one background-like event. 
The left panel corresponds to the signal-like event, 
while the right panel corresponds to the background-like event. 
In each plot, the prediction starts from the baseline value $E[f(X)]$, 
and the SHAP contributions of the input features
are then added to obtain the
final classifier output $f(x)$. 
Red bars indicate positive contributions, 
which push the event toward the signal-like region, 
while blue bars indicate negative contributions, 
which push the event toward the background-like region.
For the signal-like event in the left panel of Fig.~\ref{ML2}, 
the final classifier output is positive.
The largest positive contribution comes from $H_T^j$, 
with additional contributions from dijet invariant 
masses and $M_{\rm miss}$. 
This behaviour agrees with the expected signal 
topology, where the charged Higgs bosons decay 
through $H^\pm\to W^\pm H$, 
followed by hadronic $W$ decays and invisible 
DM particles in the final state. 
By contrast, the background-like event in the 
right panel has a negative classifier output. 
Although some features give positive contributions, 
they are outweighed by negative contributions
from other kinematic variables. 
This shows that the classifier does not rely 
on a single observable, 
but instead uses the combined event topology.
Overall, the SHAP analysis demonstrates that the XGBoost classifier distinguishes signal from background by exploiting physically meaningful kinematic observables.
The most important features are directly 
connected to $H_T^j$
the dijet invariant-mass structure, 
and the missing invariant mass expected 
in charged Higgs pair production. 
This provides a useful validation of
the ML analysis used in the main text.
\section*{Kinematic Observables}  
We present all kinematic observables
employed in this paper. The kinematic 
variables listed in this Appendix
have been implemented in 
{\tt MadGraph5\_aMC@NLO}~\cite{Alwall:2014hca}. 
First, the transverse momentum of 
particle $i$ is defined as follows:
\begin{eqnarray}
p_{T,i} = \sqrt{p_{x,i}^2 + p_{y,i}^2}.
\end{eqnarray}
The pseudorapidity of particle 
$i$ is defined as follows:
\begin{eqnarray}
\eta_i &=& \dfrac{1}{2}
\ln\!\left( \dfrac{|\vec p_i|+p_{z,i}}
{|\vec p_i|-p_{z,i}} \right).
\end{eqnarray}
The azimuthal angle is calculated
as follows:
\begin{align}
\tan\phi_i = \frac{p_{y,i}}{p_{x,i}}.
\end{align}
The invariant mass of two particles
$i$ and $j$ is defined as follows:
\begin{align}
m_{ij} &= \sqrt{
(E_i+E_j)^2 
-
(p_{x,i}+p_{x,j})^2
-
(p_{y,i}+p_{y,j})^2
-
(p_{z,i}+p_{z,j})^2}.
\end{align}
Azimuthal-angle separation two particles
is
\begin{align}
\Delta\phi_{ij} = \phi_i-\phi_j.
\end{align}
and angular distance of them is
\begin{align}
\Delta R_{ij} =
\sqrt{
\left(\eta_i-\eta_j\right)^2
+
\left(\Delta\phi_{ij}\right)^2 }.
\end{align}
The visible momentum is defined as
the sum of the momenta of all visible
final-state particles, given by:
\begin{align}
p_{\mathrm{vis}}^\mu = 
\sum_{i\in \mathrm{vis}} p_i^\mu.
\end{align}
The missing momentum is
\begin{align}
\label{pmiss}
p_{\mathrm{miss}}^\mu
\equiv
p_{\mathrm{ini}}^\mu -
p_{\mathrm{vis}}^\mu.
\end{align}
Therefore, missing transverse energy
is get 
\begin{align}
E_T^{\mathrm{miss}} =
\sqrt{ p_{x,\mathrm{miss}}^2
+
p_{y,\mathrm{miss}}^2
}.
\end{align}
The sum of jet transverse momentum is:
\begin{align}
H_T^{\mathrm{jet}} =
\sum_{i\in \mathrm{jets}} p_{T,i}.
\end{align}
Visible transverse momentum sum:
\begin{eqnarray}
S_T^{\mathrm{vis}} = 
\sum_{i\in \mathrm{vis}} p_{T,i}.
\end{eqnarray}
We also have four-jet system momentum:
\begin{align}
p_{4j}^{\mu}
&= 
p_{j_1}^{\mu}
+
p_{j_2}^{\mu}
+
p_{j_3}^{\mu}
+
p_{j_4}^{\mu}.
\end{align}
Transverse momentum of the four-jet system:
\begin{align}
p_T^{4j} = \sqrt{ p_{x,4j}^{2} + p_{y,4j}^{2} }.
\end{align}
Invariant mass of the four-jet system:
\begin{align}
m_{4j}
=
\sqrt{
E_{4j}^{2}
-
p_{x,4j}^{2}
-
p_{y,4j}^{2}
-
p_{z,4j}^{2}
}.
\end{align}
\end{document}